\documentclass[a4paper,fleqn]{cas-sc}

\usepackage[numbers,sort,compress]{natbib}

\let\printorcid\relax 
\usepackage{amsmath}
\hypersetup{
     colorlinks   = true,
     allcolors    = blue
}
\def\[#1\]{\begin{equation}#1\end{equation}}

\makeatletter
\providecommand{\doi}[1]{%
  \begingroup
    \let\bibinfo\@secondoftwo
    \urlstyle{rm}%
    \href{http://dx.doi.org/#1}{%
      doi:\discretionary{}{}{}%
      \nolinkurl{#1}%
    }%
  \endgroup
}
\makeatother

\begin{document}
\let\WriteBookmarks\relax
\def\floatpagepagefraction{1}
\def\textpagefraction{.001}
\shorttitle{Isotropic finite-difference approximations for phase-field simulations of polycrystalline alloy solidification}
\shortauthors{K. Ji et~al.}

\title [mode = title]{Isotropic finite-difference approximations for phase-field simulations of polycrystalline alloy solidification}                     



\author[1]{Kaihua Ji}
\author[1]{Amirhossein Molavi Tabrizi}
\author[1]{Alain Karma\corref{cor1}}

\cortext[cor1]{Corresponding author: a.karma@northeastern.edu}

\address[1]{Physics Department and Center for Interdisciplinary Research on Complex Systems, Northeastern University, Boston, Massachusetts 02115}

\begin{abstract}
Phase-field models of microstructural pattern formation during alloy solidification are commonly solved numerically using the finite-difference method, which is ideally suited to carry out computationally efficient simulations on massively parallel computer architectures such as Graphic Processing Units. However, one known drawback of this method is that the discretization of differential terms involving spatial derivatives introduces a spurious lattice anisotropy that can influence the solid-liquid interface dynamics. We find that this influence is significant for the case of polycrystalline dendritic solidification, where the crystal axes of different grains do not generally coincide with the reference axes of the finite-difference lattice. In particular, we find that with the commonly used finite-difference implementation of the quantitative phase-field model of binary alloy solidification, both the operating state of the dendrite tip and the dendrite growth orientation are strongly affected by the lattice anisotropy. To circumvent this problem, we use known methods in both real and Fourier space to derive finite-difference approximations of leading differential terms in 2D and 3D that are isotropic at order $h^2$ of the lattice spacing $h$. Importantly, those terms include the divergence of the anti-trapping current that is found to have a critical influence on pattern selection. The 2D and 3D discretizations use an approximated form of the anti-trapping current that facilitates the Fourier-space derivation of the associated isotropic differential operator at $O(h^2)$, but we also derive a 2D discretization of the standard form of this current. Finally, we present 2D and 3D phase-field simulations of alloy solidification, showing that the isotropic finite-difference implementations dramatically reduce spurious lattice anisotropy effects, yielding both the tip operating state and growth direction of the dendrite that are nearly independent of the angle between the crystal and finite-difference lattice axes.
\end{abstract}




\maketitle

\section{Introduction}

The phase-field (PF) method is now established as a method of choice to simulate microstructural pattern formation during alloy solidification \cite{Boettinger2000,chen2002phase,boettinger2002phase,asta2009solidification,Karma2016}. 
In many applications, the partial differential equations (PDEs) of phase-field models are implemented using a finite-difference discretization of spatial derivatives. Such an implementation is ideally suited to carry out computationally efficient simulations on massively parallel computer architectures such as Graphic Processing Units (GPUs) \cite{Yamanaka2011,Tourret2015,Tourret2017,Clarke2017}, which are making it increasingly feasible to simulate solidification patterns on experimentally relevant length and time scales \cite{Clarke2017,Shimokawabe,Mota2021}.
One drawback of the finite-difference method, however, is that it introduces a spurious anisotropy that can influence the interface dynamics when the equations are discretized on a regular mesh with mesh points arranged on a square or cubic lattice in 2D or 3D, respectively. This is especially true for dendritic microstructures in metallic alloys owing to the property that the dendrite growth direction and the tip operating state are controlled by the weak anisotropy of the excess free-energy of the solid-liquid interface \cite{Langer1989,Barbieri1989,BenAmar1993}. Even though the lattice anisotropy can in principle be made arbitrarily small by reducing the lattice spacing (mesh size) $h$, it is not negligibly small in typical simulations where $h$ is comparable to the width of the spatially diffuse interface. For simulations of a single grain where the crystal and the finite-difference mesh have the same symmetry (e.g., cubic symmetry), and the principal crystal axes coincides with those of the mesh (e.g., the $\langle 100 \rangle $ crystal axes that set the $\langle 100 \rangle $ dendrite growth directions), it is possible to compute an effective cubic anisotropy that results from the combination of the lattice anisotropy and the anisotropy incorporated in the phase-field model  \cite{Karma1998}. However, this is no longer possible for polycrystalline solidification since the principal crystal axes of grains of different orientations do not typically coincide with the principal axes of the underlying cubic mesh.  This is also true for a single grain when the crystal and the mesh have different symmetries (e.g., in using a cubic mesh to model dendritic solidification in alloy systems with hexagonal symmetry).

To date, quantitative phase-field simulations of polycrystalline alloy solidification \cite{Tourret2015,Tourret2017,Clarke2017,Pineau2018} have been typically carried out using standard finite-difference schemes where, with the exception of Laplacian terms, the discretization of spatial derivative terms are not rotationally invariant at order $h^2$ of the lattice spacing $h$. However, in a recent study of the growth competition of columnar grains focusing on a fully developed dendritic regime \cite{Dorari2021}, where the entire dendrite tip region grows under isothermal conditions, we found that those schemes are insufficient to accurately resolve the dendrite growth orientation and the dendrite tip operating state for grains that have a large misorientation with respect to the principal axes of the lattice. The lattice anisotropy causes a significant deviation of the dendrite growth orientation from the principal crystal axes and also affects the tip selection parameter $\sigma^*=2 D d_0^*/\rho^2V$ determined by microscopic solvability theory  \cite{Langer1989,Barbieri1989,BenAmar1993} (where $\rho$ and $V$ are the tip velocity and radius, respectively, $D$ is the solute diffusivity, and $d_0^*$ is the capillary length). To overcome this difficulty, we developed a finite-difference scheme where all relevant spatial derivative terms influencing the interface dynamics are rotationally invariant at order $h^2$. This new scheme was used to carry out the study of grain competition reported in Ref. \cite{Dorari2021}. The present paper provides a detailed exposition of the theoretical development of this scheme and numerical tests for isothermal and directional solidification in 2D and 3D. The development is carried out for a quantitative model of alloy solidification \cite{Karma2001,Echebarria2004} by considering the solidification of a single grain with cubic symmetry where the crystal axes have an arbitrary orientation with respect to the principal axes of the cubic finite-difference lattice. The new scheme, however, should be broadly applicable to a wide range of PF models of polycrystalline solidification. Those include the formulation employed to study grain competition in Ref. \cite{Dorari2021}, which makes use of an indicator field \cite{Tourret2015,Tourret2017,Clarke2017,Pineau2018} to distinguish different grains but does not describe grain-grain interactions that occur at the late stages of solidification. They also include various formulations that use an orientation field \cite{Kobayashi1998,Granasy2002,Warren2003,Granasy2004,Pusztai2005,Kobayashi2005,Tourret2015}, multiple order parameters \cite{Steinbach1996,Fan1997,Garcke2006,Steinbach1999,Folch2005,Moelans2008}, or a vector field \cite{Pinomaa2021} to model those interactions. While the formulations that resolve grain-grain interactions contain additional spatial derivative terms than those considered here, which are linked to extra fields that describe orientational order, those terms mainly affect grain boundary properties. Since grain boundaries tend to be more anisotropic than the solid-interface energy, it may not be necessary to use higher-order rotationally-invariant discretizations to quantitatively resolve their behavior. The present paper focuses primarily on suppressing the effect of the lattice anisotropy on the solidification dynamics of individual grains of arbitrary orientation for modeling the formation of well-developed dendritic alloy microstructures \cite{Dorari2021}. Grain coalescence at the late stages of solidification is not considered.
 
Let us briefly outline the various spatial derivative terms that need to be considered for this purpose. For the PF model of dendritic crystal growth in a pure melt \cite{Karma1996,Karma1998,Glasner2001}, the Laplacian of a PF and a thermal field are the leading differential terms whose finite-difference implementations can have a major influence on PF simulations. In 2D, the well-known isotropic discretization of Laplacian utilizes a nine-point stencil that contains four nearest-neighbor points in $\left<10\right>$ directions and four next-nearest-neighbor points in $\left<11\right>$ directions on a square lattice. As will be briefly reviewed in Section \ref{Sec:lap_real_2D} of this paper, the isotropic discretization of 2D Laplacian has a leading error that is rotationally invariant at order $h^2$. In 3D, a commonly used isotropic discretization of the Laplacian utilizes 18 lattice points, including six nearest-neighbor points in $\left<100\right>$ directions and twelve next-nearest-neighbor points in $\left<110\right>$ directions on a cubic lattice, which is found to be sufficient for the accurate simulation of capillary and kinetic anisotropies \cite{Glasner2001}. Meanwhile, several other forms of 3D isotropic Laplacian that utilize additional lattice points in $\left<111\right>$ directions have been proposed in previous studies \cite{Kumar2004,Patra2006,Thampi2013}. We will also discuss and give remarks on those known isotropic discretizations of 3D Laplacian in Section \ref{Sec:3D_1}. 

In contrast, for the quantitative PF model of alloy solidification \cite{Karma2001,Echebarria2004}, we need to consider the isotropic discretization of additional differential terms that have more complex forms than the Laplacian. The continuity equation that describes the solute transport in the mesoscale interface region contains the divergence of a solutal flux, which consists of the gradient of the chemical potential and the anti-trapping current $\vec j_{at}$. The quantitative anti-trapping current ensures the limit of local thermodynamic equilibrium at the solid-liquid interface \cite{Karma2001}. Hence, the PF model of alloy solidification remains quantitative while a mesoscopic interface thickness $W$ is used to simulate the interface evolution on experimental length and time scales. Since the amount of the compensating anti-trapping current increases with the velocity and thickness of a diffuse interface, the discretization of the anti-trapping term (divergence of $\vec j_{at}$) can have a critical influence on PF simulations, especially when the interface velocity is large and/or a large $W$ is chosen. In the PF model, the divergences of the chemical potential gradient and $\vec j_{at}$ take generalized forms $\vec{\nabla} \cdot ( \alpha {\vec{\nabla}\beta} )$ and $\vec{\nabla} \cdot ( \alpha {\vec{\nabla}\beta}/{|\vec{\nabla}\beta|} )$, respectively, where $\alpha$ and $\beta$ can be arbitrary scalar fields. While the isotropic discretization of the former can be achieved at order $h^2$ in a straightforward manner as introduced in Section \ref{Sec:2D_Div_dis} (2D) and Section \ref{Sec:3D_Div_dis} (3D), the isotropic discretization of the latter is more complex. In 2D, we derive the isotropic discretization of $\vec{\nabla} \cdot ( \alpha {\vec{\nabla}\beta}/{|\vec{\nabla}\beta|} )$ by using an extended stencil that contains 21 lattice points (Appendix \ref{Sec:appendix_vanish}). But a similar strategy in 3D becomes extremely inefficient for numerical implementations. Therefore, we propose an approximation to the anti-trapping current by simply assuming that, in the interface region, the PF profile remains close to the standard tangent hyperbolic profile corresponding to a stationary equilibrium interface, so that the divergence of the approximated anti-trapping current can be written in the form $\vec{\nabla} \cdot ( \alpha {\vec{\nabla}\beta} )$. A comparison of 2D PF simulations using the approximated anti-trapping current with an isotropic discretization and the full anti-trapping current discretized with a larger stencil shows that this approximation is quantitatively accurate and turns out to be modestly more computationally efficient.

Next, we examine the rotational invariance of finite-difference schemes that contain spatial discretizations of all leading differential terms in the real or Fourier space. We denote a group of lattice points in a set of equivalent directions as a ``lattice shell''. In 2D, the isotropic scheme combines two base discretizations that utilize lattice shells in the $\left<10\right>$ and $\left<11\right>$ directions; In 3D, the isotropic scheme combines two or three of the base discretizations that utilize lattice shells in the $\left<100\right>$, $\left<110\right>$, and $\left<111\right>$ directions. Previous studies proposed several isotropic discretizations of 3D Laplacian individually \cite{Kumar2004,Patra2006,Thampi2013}, but a systematic investigation is still lacking. We show that, there are various combinations of base discretizations to obtain an isotropic 3D Laplacian at order $h^2$, and only one combination among them utilizes the first two bases with lattice shells in the $\left<100\right>$ and $\left<110\right>$ directions (the 18-point formula). The same conclusion applies to the other leading differential terms in 3D. Finally, we present 2D and 3D PF simulations of isothermal and directional solidification with different finite-difference schemes, showing that isotopic schemes can significant improve the rotational invariance and are necessary for PF simulations of solidification in the well-developed dendritic regime. 

The paper is organized as follows. Section \ref{Sec:PF_model} introduces the PF model, identifies the leading differential terms, and proposes an approximation to the anti-trapping current. Section \ref{Sec:2D_1} introduces different discretizations in 2D. Section \ref{Sec:2D_2} examines the rotational invariance of discretizations in both real and Fourier space and then implements several finite-difference schemes for 2D PF simulations of alloy solidification. Section \ref{Sec:3D_1} introduces different discretizations in 3D. Section \ref{Sec:3D_2} examines the rotational invariance of discretizations in the 3D Fourier space and then implements several finite-difference schemes for 3D PF simulations of alloy solidification. The isotropic discretizations of the gradient and the standard anti-trapping that requires an extended stencil are discussed in Appendices.

\section{Phase-field model} \label{Sec:PF_model}
We consider a well-established quantitative PF model for solidification of dilute binary alloys \cite{Karma2001,Echebarria2004}. The model can be applied to different thermal conditions, including the isothermal condition for free dendrite growth \cite{Karma2001} and the frozen temperature approximation for directional solidification \cite{Echebarria2004}. The model utilizes the thin-interface asymptotics \cite{Karma1998,Echebarria2004} to make the diffuse interface width $W$ much larger than the capillarity length $d_0$ while remain quantitative. An anti-trapping current is introduced into the model to compensate for spurious solute trapping due to a large $W$ \cite{Karma2001}. The conventional PF $\varphi$ is a scalar that takes on constant values in solid $(+1)$ and liquid $(-1)$ and vary smoothly across the diffuse interface. Here we use a nonlinear preconditioning PF $\psi$ defined by
\[
\varphi(\mathbf{r}, t)=\tanh \left\{\frac{\psi(\mathbf{r}, t)}{\sqrt{2}}\right\}
\]
in order to enhance the numerical stability at larger grid spacings \cite{Glasner2001}. This preconditioning PF brings an additional gradient term that needs to be isotropically discretized, as will be discussed in Section \ref{Sec:leading_diff}. In addition, there is a dimensionless supersaturation $U$ field coupling to the $\psi$ field.

\subsection{Evolution equations} \label{Sec:evolution_eqs}
The evolution of the preconditioned PF $\psi$ and the dimensionless supersaturation $U$ follows partial differential equations:
\[
\begin{split} \label{PF_a}
\widetilde{F}_1 a_s(\mathbf{n})^2 \frac{\partial \psi}{\partial t} =& a_s(\mathbf{n})^2 \left( \nabla^2 \psi - \varphi \sqrt{2} |\vec{\nabla} \psi|^2 \right) + \vec{\nabla} \left[ a_s(\mathbf{n})^2 \right] \cdot \vec{\nabla} \psi + \sum_{q}\left[\partial_{q}\left(|\vec{\nabla} \psi|^{2} a_{s}(\mathbf{n}) \frac{\partial a_{s}(\mathbf{n})}{\partial\left(\partial_{q} \psi\right)}\right)\right] \\
&+ \sqrt{2} \left[ \varphi -\lambda(1-\varphi^2) \widetilde{F}_2 \right],
\end{split}
\]
and
\[
\begin{split} \label{PF_b}
(1+k-(1-k) \varphi) \frac{\partial U}{\partial t}=&\widetilde{D} \vec{\nabla} \cdot[(1-\varphi) \vec{\nabla} U] + \vec{\nabla} \cdot\left[(1+(1-k) U) \frac{(1-\varphi^2)}{2} \frac{\partial \psi}{\partial t} \frac{\vec{\nabla} \psi}{|\vec{\nabla} \psi|}\right] \\
&+[1+(1-k) U] \frac{\left(1-\varphi^{2}\right)}{\sqrt{2}} \frac{\partial \psi}{\partial t},
\end{split}
\]
with
\[
U=\frac{1}{1-k}\left[\frac{c / c_{l}^{0}}{(1-\varphi) / 2+k(1+\varphi) / 2}-1\right], \label{Supersaturation}
\]
where $c$ is a field of the solute concentration in molar fraction, $c^0_l=(T_m-T'_0)/|m|$ is the liquidus concentration at the reference temperature $T=T'_0$, with $T_m$ the melting temperature of pure solvent and $m$ the slope of liquidus. $k$ is the interface solute partition coefficient, and we use $k=0.1$ in simulations throughout this paper. The index $q$ sums over the spatial coordinates, i.e., $q=x,y$ in 2D and $q=x,y,z$ in 3D. The length and time are scaled by the diffuse interface thickness $W$ and the relaxation time $\tau_0$ \cite{Karma2001,Echebarria2004}, respectively. The dimensionless diffusivity $\widetilde{D}$ is given by
\[
\widetilde{D}=\frac{D \tau_{0}}{W^{2}}=a_{1} a_{2} \frac{W}{d_{0}},
\]
with $a_1=5\sqrt{2}/8$ and $a_2=47/75$ \cite{Karma1998}. The capillary length $d_0$ is defined at the reference temperature $T_0$ by
\[
d_0=\frac{\Gamma}{\Delta T_0}, \label{d_0}
\]
where $\Gamma$ is the Gibbs-Thomson coefficient of the solid-liquid interface, and $\Delta T_0=|m|(1-k)c^0_l$ is the freezing range. The coupling factor in this model is $\lambda=a_1 W/d_0$.

For directional solidification, the reference temperature is chosen to be $T'_0=T_0$ such that $c^0_l=(T_m-T_0)/|m|=c_{\infty}/k$, where $c_{\infty}$ is the nominal concentration. We assume the temperature field is frozen with an initial temperature gradient $G$. Then the temperature $T$ at a position $x$ aligned with the pulling velocity $V_p$ is given by
\[
T=T_0+G(x-V_p t).
\]
Two thermal functions in Eq. \eqref{PF_a} are
\[
\widetilde{F}_{1}(x, t)=\left[1-(1-k) \frac{x-\widetilde{V}_p t}{\tilde{l}_{T}}\right], \label{Thermal_F1}
\]
and
\[
\widetilde{F}_{2}(x, t)=\left[U+\frac{x-\widetilde{V}_p t}{\tilde{l}_{T}}\right]. \label{Thermal_F2}
\]
The dimensionless pulling velocity $\widetilde{V}_p$ and thermal length $\tilde{l}_{T}$ are defined by
\[
\widetilde{V}_{p}=\frac{V_{p} \tau_{0}}{W}=a_{1} a_{2} \frac{V_{p} d_{0}}{D}\left(\frac{W}{d_{0}}\right)^{2},
\]
and
\[
\widetilde{l}_{T}=\frac{l_{T}}{W},
\]
where $l_{T}={|m|(1-k) c_{l}^{0}}/{G}$ is the dimensional thermal length. Note that the length and time in the thermal equations are in the units of $W$ and $\tau_0$, respectively. 

For the isothermal growth condition, the reference temperature $T'_0$ is larger than $T_0$, and the definition of the capillary length becomes different. With a supersaturation $\Omega=(c^0_l-c_{\infty})/[c^0_l(1-k)]$, there is a relation
\[
d_0^*=\frac{1-(1-k)\Omega}{k} d_0.
\]
Note that $d_0^*$ reduces to $d_0$ in the limit of $\Omega=0$. Meanwhile, the two thermal functions in Eq. \eqref{PF_a} are simply $\widetilde{F}_1=1$ and $\widetilde{F}_2=U$. They can be inferred from Eqs. \eqref{Thermal_F1}-\eqref{Thermal_F2} by considering the isothermal limits, i.e., $G \to 0$ and $l_T \to \infty$.

We consider a cubic symmetry of the interface anisotropy $\gamma = \gamma_0 ( 1+\epsilon_4 \cos{4\theta})$, where $\gamma_0$ is the average interface free energy in a $\left<10\right>$ or $\left<100\right>$ plane, $\epsilon_4$ is a measure of the strength of the anisotropy, and $\theta$ is the angle between the local surface normal vector and a fixed crystalline axis. For a crystal orientation angle $\alpha_0$ with respect to the reference axis, the anisotropy function in the PF model is
\[
a_s(\theta)=1+\epsilon_4 \cos{[4(\theta-\alpha_0)]}. \label{anis_function}
\]
We impose the angle $\alpha_0$ in PF simulations using the rotation matrix methods in 2D \cite{Tourret2015} and 3D \cite{Tourret2017}.

\subsection{Leading differential terms} \label{Sec:leading_diff}
We develop isotropic discretizations for all the spatial derivatives in the evolution Eqs. \eqref{PF_a}-\eqref{PF_b} except for the anisotropy-related terms in Eq. \eqref{PF_a} that scale linearly with $\epsilon_4$ in the $\epsilon_4 \ll 1$ limit. This is because the anisotropy of the excess free-energy of the solid-liquid interface ($\epsilon_4$) is typically small in metallic systems and transparent organic analogs in Eq. \eqref{anis_function}. Therefore it suffices to use a standard finite-difference discretization for those terms since the effective anisotropy introduced by this choice scales as the product of $\epsilon_4$ and the grid-induced anisotropy, which is much smaller than $\epsilon_4$ itself and has a negligible effect 
on the dendrite growth direction and tip operating state in PF simulations. In contrast, for all other spatial derivatives including the anti-trapping current in Eq. \eqref{PF_b}, isotropic discretizations need to be used to accurately resolve those quantities.

For evolution of the $\psi$ field in Eq. \eqref{PF_a}, leading differential terms include the Laplacian term
\[
\mathcal{L}(\mathbf{r})=\nabla^2 \psi,
\]
and the gradient square term
\[
\mathcal{G}(\mathbf{r})=|\vec{\nabla} \psi|^2.
\]
Note that for a PF model with the $\varphi$ field (Eq. (132) of Ref. \cite{Echebarria2004}), $\mathcal{G}(\mathbf{r})$ is not one of the leading differential terms anymore. We will address the isotropic discretization of $\mathcal{L}(\mathbf{r})$ in the main text and discuss the discretization of $\mathcal{G}(\mathbf{r})$ in Appendix \ref{Sec:dis_grad}. Although the isotropic finite-difference implementation of $\mathcal{L}(\mathbf{r})$ is found to be sufficient for PF simulations of solidification of a pure undercooled melt \cite{Karma2001}, it is insufficient for PF simulations of alloy solidification when a solutal field is coupled to the PF.

For evolution of the $U$ field in Eq. \eqref{PF_b}, there are two spatially differential terms, i.e., the divergences of the supersaturation gradient and the anti-trapping current. The former is denoted by
\[
\mathcal{D}(\mathbf{r})=\vec{\nabla} \cdot[(1-\varphi) \vec{\nabla} U], \label{D_term}
\]
and the latter is denoted by
\[
\mathcal{A}(\mathbf{r})=\vec{\nabla} \cdot\left[(1+(1-k) U) \frac{(1-\varphi^2)}{2} \frac{\partial \psi}{\partial t} \frac{\vec{\nabla} \psi}{|\vec{\nabla} \psi|}\right].
\]
Note that the normal vector has two analytically equivalent forms, i.e., $\hat{\mathbf{n}} \equiv {\vec{\nabla} \psi}/{|\vec{\nabla} \psi|} \equiv {\vec{\nabla} \varphi}/{|\vec{\nabla} \varphi|}$. Thus, the other expressions of $\mathcal{A}(\mathbf{r})$ can be inferred by knowing the relation ${\sqrt{2} {\partial_t \varphi}=(1-\varphi^2) \partial_t \psi}$, which gives
\[
\mathcal{A}(\mathbf{r})= \vec{\nabla} \cdot\left[(1+(1-k) U) \frac{1}{\sqrt{2}} \frac{\partial \varphi}{\partial t} \frac{\vec{\nabla} \varphi}{|\vec{\nabla} \varphi|}\right]. \label{Anti_convention}
\]
Although isotropic discretizations of both $\mathcal{D}(\mathbf{r})$ and $\mathcal{A}(\mathbf{r})$ will be addressed in the main text, the latter can be dominant when the solid-liquid interface is far from equilibrium, or when a large $W$ is chosen in the PF simulation.

\subsection{Approximated anti-trapping current}
Those two differential terms $\mathcal{D}(\mathbf{r})$ and $\mathcal{A}(\mathbf{r})$ in the evolution equation of the $U$ field share a similar form, i.e., the divergence of a spatially differential term. As we will show in Section \ref{Sec:2D_Div_dis} and Section \ref{Sec:3D_Div_dis}, the isotropic discretizations of $\mathcal{D}(\mathbf{r})$ can be achieved in 2D and 3D by utilizing only two lattice shells. However, the isotropic discretizations of $\mathcal{A}(\mathbf{r})$ require extended stencils as discussed in Appendix \ref{Sec:appendix_vanish}, which would be inefficient for numerical implementations especially in 3D. Thus, we hope to eliminate $|\vec{\nabla} \psi|$ at the denominator of the anti-trapping current, and transform $\mathcal{A}(\mathbf{r})$ into a same generalized form of $\mathcal{D}(\mathbf{r})$, i.e., $\vec{\nabla} \cdot ( \alpha {\vec{\nabla}\beta} )$.

The one-dimensional stationary solution of the PF has a standard tangent hyperbolic profile $\varphi_0(x)=-\tanh{ ( {x}/{\sqrt{2}} ) }$, where $x$ is the dimensionless spatial coordinate scaled by $W$. The derivative of this stationary solution is $\partial_x \varphi_0=-(1-\varphi_0^2)/\sqrt{2}$, which does not contain spatial derivatives on the right-hand side. Thus, we can extend it into 2D and 3D in terms of the gradient, i.e.,
\[
|\vec{\nabla} \varphi_0 | = \frac{(1-\varphi_0^2)}{\sqrt{2}}.
\]
Replacing $|\vec{\nabla} \varphi |$ in Eq. \eqref{Anti_convention} by $|\vec{\nabla} \varphi_0 |$, we can approximate $\mathcal{A}(\mathbf{r})$ by \begin{align}
\bar{\mathcal{A}}(\mathbf{r})=& \frac{1}{\sqrt{2}} \vec{\nabla} \cdot\left[(1+(1-k) U) \frac{\partial \varphi}{\partial t} {\vec{\nabla} \psi} \right] \label{Anti_approx} \\
=& \frac{1}{\sqrt{2}} \vec{\nabla} \cdot\left[(1+(1-k) U) \frac{\partial \psi}{\partial t} {\vec{\nabla} \varphi} \right], \label{Anti_approx2}    
\end{align}
where a bar is used to denote the approximation and to distinguish it from $\mathcal{A}(\mathbf{r})$. The above two forms of $\bar{\mathcal{A}}(\mathbf{r})$ are analytically equivalent, and we consider only the form in Eq. \eqref{Anti_approx} in the following. By adopting this approximation, the generalized form of the anti-trapping changes from $\mathcal{A}(\mathbf{r})=\vec{\nabla} \cdot ( \alpha_1 {\vec{\nabla}\psi}/{|\vec{\nabla}\psi|} )$ to $\bar{\mathcal{A}}(\mathbf{r})=\vec{\nabla} \cdot ( \alpha_2 {\vec{\nabla}\psi} )$, where $\alpha_1$ and $\alpha_2$ are two scalars, so that it can be discretized isotropically in a more concise manner similar to $\mathcal{D}(\mathbf{r})$. The numerical benefit of using $\bar{\mathcal{A}}(\mathbf{r})$ will be further discussed in Section \ref{Sec:2D_results}.

\section{Isotropic discretizations in 2D} \label{Sec:2D_1} 
\begin{figure}[h]
\centering
\includegraphics[scale=0.65]{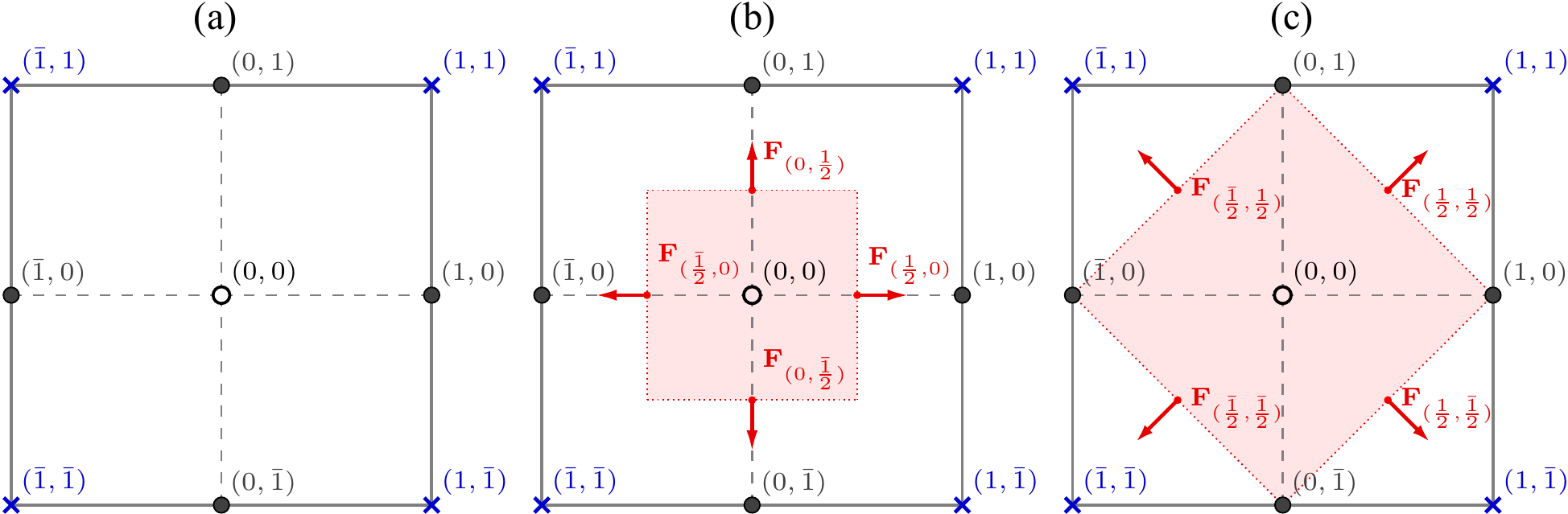}
\caption{(a) A unit cell in the 2D square lattice. The origin point is marked by a black hallow circle. The $\left<10\right>$ lattice shell is marked by black filled circles and the $\left<11\right>$ lattice shell is marked by blue crosses. (b) The fluxes in the $\left<10\right>$ directions for evaluating a divergence. (c) The fluxes in the $\left<11\right>$ directions for evaluating a divergence. \label{fig:2D_Fluxes}}
\end{figure}

We first introduce a notation system that can be used to describe the discretization of all leading differential terms in 2D. Consider an arbitrary differential term $\mathcal{T}(\mathbf{r})$. It is discretized by finite difference on a 2D square lattice with the lattice spacing $\Delta x=\Delta y=h$, as shown in Fig. \ref{fig:2D_Fluxes}(a). The coordinates of a lattice point $(r_1 h,r_2 h)$ are abbreviated as $(r_1,r_2)$. We use $\mathcal{T}_{1,0}(\mathbf{r})$ to denote the discretization involving only the $\left<10\right>$ lattice shell, and $\mathcal{T}_{0,1}(\mathbf{r})$ to denote the discretization involving only the $\left<11\right>$ lattice shell. Note that the origin point $(0,0)$ is involved by default in both cases. Since each lattice shell can be used independently to discretize $\mathcal{T}(\mathbf{r})$, we call $\mathcal{T}_{1,0}(\mathbf{r})$ and $\mathcal{T}_{0,1}(\mathbf{r})$ two base discretizations. These two bases can be combined to produce a complete set of discretizations on a nine-point stencil. Thus, we can use the notation $\mathcal{T}_{s_1,s_2}(\mathbf{r})$, or simply $\mathcal{T}_{s_1,s_2}$ to denote an arbitrary discretization with subscript indicating the ratio of two bases, such that 
\[
\mathcal{T}_{s_1,s_2}=\frac{s_1}{s_1+s_2} \mathcal{T}_{1,0} + \frac{s_2}{s_1+s_2} \mathcal{T}_{0,1}. \label{tau_2D}
\]
Note that the sum of weights on two bases always equals to 1. In the following, the same notations are applied to all differential terms, including $\mathcal{L}(\mathbf{r})$, $\mathcal{G}(\mathbf{r})$, $\mathcal{D}(\mathbf{r})$, $\mathcal{A}(\mathbf{r})$, and $\bar{\mathcal{A}}(\mathbf{r})$.

\subsection{Laplacian}
In the continuum limit, the Laplacian of a scalar field $\psi(\mathbf{r})$ is defined by $\mathcal{L}(\mathbf{r})=\nabla^2 \psi(\mathbf{r})$. The scalar field has a specific value $\psi(\mathbf{r}+\mathbf{c})$ at the lattice point $(\mathbf{r}+\mathbf{c})$, which is abbreviated by $\psi_{(r_1,r_2)}$ with $\mathbf{r}=0$ and $\mathbf{c}=(r_1 h \hat{\mathbf{x}}+r_2 h \hat{\mathbf{y}})$. The most commonly used discrete Laplacian employs a five-point stencil, or the $\left<10\right>$ lattice shell. It is referred to as $\mathcal{L}_{1,0}$ in this paper, and has the form
\[
\mathcal{L}_{1,0}=\frac{1}{h^2} \left[ \sum_{i=1}^4 \psi_{\left<10\right>}^i-4 \psi_{\left(0,0\right)} \right], \label{2D_L10}
\]
where the index $i$ sums over values of $\psi$ in the $\left<10\right>$ lattice shell, i.e., $\psi_{\left<10\right>}^i$. Here $\mathcal{L}_{1,0}$ constitutes one base for discretizing $\mathcal{L}(\mathbf{r})$. Another base $\mathcal{L}_{0,1}$ employs the $\left<11\right>$ lattice shell, and has the form
\[
\mathcal{L}_{0,1}=\frac{1}{2h^2} \left[ \sum_{i=1}^4 \psi_{\left<01\right>}^i-4 \psi_{\left(0,0\right)} \right], \label{2D_L01}
\]
where the index $i$ sums over values of $\psi$ in the $\left<01\right>$ lattice shell, i.e., $\psi_{\left<01\right>}^i$. The $\left<11\right>$ lattice shell itself is equivalent to a regular square lattice with the grid spacing $\sqrt{2}h$. This is the reason why the coefficient is $1/h^2$ in Eq. \eqref{2D_L10}, and $1/2h^2$ in Eq. \eqref{2D_L01}.

The isotropic Laplacian employs a nine-point stencil, or $\left<10\right>$ and $\left<01\right>$ lattice shells. It gives $2/3$ weight to $\mathcal{L}_{1,0}$ and $1/3$ weight to $\mathcal{L}_{0,1}$. This isotropic Laplacian is referred to as $\mathcal{L}_{2,1}$ following the notation defined in Eq. \eqref{tau_2D}, and has the form
\[
\mathcal{L}_{2,1}= \left( \frac{2}{3} \mathcal{L}_{1,0}+\frac{1}{3}\mathcal{L}_{0,1} \right)=
\frac{1}{6h^2} \left[ 4\sum_{i=1}^4 \psi_{\left<10\right>}^i + \sum_{i=1}^4 \psi_{\left<01\right>}^i -20 \psi_{\left(0,0\right)} \right] \label{2D_L21}.
\]

\subsection{Divergence} \label{Sec:2D_Div_dis}
We consider a generalized divergence of the form $\widetilde{\mathcal{D}}(\mathbf{r})=\vec{\nabla} \cdot \mathbf{F}=\vec{\nabla} \cdot ( \alpha {\vec{\nabla}\beta} )$, where $\mathbf{F}=(\alpha \vec{\nabla}\beta) $ is a flux, $\alpha$ and $\beta$ are two scalar fields. In the following, we simply call $\widetilde{\mathcal{D}}(\mathbf{r})$ a divergence. The same discretization of $\widetilde{\mathcal{D}}(\mathbf{r})$ can be applied to $\mathcal{D}(\mathbf{r})$ and $\bar{\mathcal{A}}(\mathbf{r})$ in the PF model. For the diffusion term $\mathcal{D}(\mathbf{r})$ in Eq. \eqref{D_term}, two scalar fields are $\alpha=(1-\varphi)$ and $\beta=U$. For the approximated anti-trapping term $\bar{\mathcal{A}}(\mathbf{r})$ in Eq. \eqref{Anti_approx}, two scalar fields are $\alpha=[(1+(1-k) U) \partial_t \varphi]/\sqrt{2}$ and $\beta=\psi$.

A commonly used discretization of $\widetilde{\mathcal{D}}(\mathbf{r})$ utilizes fluxes in the $\langle 10 \rangle$ directions \cite{Tourret2015}, as shown in Fig. \ref{fig:2D_Fluxes}(b). It constitutes a base discretization denoted by $\widetilde{\mathcal{D}}_{1,0}$ and has the form  
\[
\widetilde{\mathcal{D}}_{1,0} = \frac{1}{h} \left [ \mathbf{F}_{(1/2,0)}+\mathbf{F}_{(\bar{1}/2,0)}+\mathbf{F}_{(0,1/2)}+\mathbf{F}_{(0,\bar{1}/2)} \right], \label{divergence_10}
\]
where $\mathbf{F}_{(i,j)}$ is the flux at the off-lattice position $(i,j)$, and the flux pointing out of the unit cell is positive. This form ensures mass conservation by the virtue of symmetry. However, there can be different ways to evaluate $\mathbf{F}_{(i,j)}$ depending on how $\alpha_{(i,j)}$ at the off-lattice position $(i,j)$ is taken, which can significantly affect the rotational invariance of a discretized $\widetilde{\mathcal{D}}(\mathbf{r})$. In order to obtain an isotropic discretization of $\widetilde{\mathcal{D}}(\mathbf{r})$ later, we propose to evaluate each flux in the $\langle 10 \rangle$ directions as
\[
\begin{split}
&\mathbf{F}_{(1/2,0)} = \frac{1}{4h} \left[ \alpha_{(1,0)}+\alpha_{(0,0)}+\bar{\alpha}_{(1/2,1/2)}+\bar{\alpha}_{(1/2,\bar{1}/2)} \right] \left[ \beta_{(1,0)}-\beta_{(0,0)} \right], \\
&\mathbf{F}_{(0,1/2)} = \frac{1}{4h} \left[ \alpha_{(0,1)}+\alpha_{(0,0)}+\bar{\alpha}_{(1/2,1/2)}+\bar{\alpha}_{(\bar{1}/2,1/2)} \right] \left[ \beta_{(0,1)}-\beta_{(0,0)} \right], \\
&\mathbf{F}_{(\bar{1}/2,0)} = \frac{1}{4h} \left[ \alpha_{(0,0)}+\alpha_{(\bar{1},0)}+\bar{\alpha}_{(\bar{1}/2,1/2)}+\bar{\alpha}_{(\bar{1}/2,\bar{1}/2)} \right] \left[ \beta_{(\bar{1},0)}-\beta_{(0,0)} \right], \\
&\mathbf{F}_{(0,\bar{1}/2)} = \frac{1}{4h} \left[ \alpha_{(0,0)}+\alpha_{(0,\bar{1})}+\bar{\alpha}_{(1/2,\bar{1}/2)}+\bar{\alpha}_{(\bar{1}/2,\bar{1}/2)} \right] \left[ \beta_{(0,\bar{1})}-\beta_{(0,0)} \right],
\end{split} \label{flux_10)}
\]
where the value of $\bar{\alpha}_{(i,j)}$ at the off-lattice position $(i,j)$ is taken by averaging its four nearest neighbors. For example, $\bar{\alpha}_{(1/2,1/2)}$ equals to $[\alpha_{(1,1)}+\alpha_{(1,0)}+\alpha_{(0,1)}+\alpha_{(0,0)}]/4$ at the off-lattice position $(1/2,1/2)$.

Another base discretization $\widetilde{\mathcal{D}}_{0,1}$ utilizes fluxes in the $\left<11 \right>$ directions, as shown in Fig. \ref{fig:2D_Fluxes}(c). It has the form  
\[
\widetilde{\mathcal{D}}_{0,1} = \frac{1}{\sqrt{2} h} \left [ \mathbf{F}_{(1/2,1/2)}+\mathbf{F}_{(\bar{1}/2,1/2)}+\mathbf{F}_{(1/2,\bar{1}/2)}+\mathbf{F}_{(\bar{1}/2,\bar{1}/2)} \right], \label{divergence_11}
\]
where each flux is evaluated by 
\[
\begin{split}
&\mathbf{F}_{(1/2,1/2)} = \frac{1}{4\sqrt{2}h} \left[ \alpha_{(1,1)}+\alpha_{(0,0)}+\alpha_{(0,1)}+\alpha_{(1,0)} \right] \left[ \beta_{(1,1)}-\beta_{(0,0)} \right], \\
&\mathbf{F}_{(\bar{1}/2,1/2)} = \frac{1}{4\sqrt{2}h} \left[ \alpha_{(\bar{1},1)}+\alpha_{(0,0)}+\alpha_{(0,1)}+\alpha_{(\bar{1},0)} \right] \left[ \beta_{(\bar{1},1)}-\beta_{(0,0)} \right], \\
&\mathbf{F}_{(1/2,\bar{1}/2)} = \frac{1}{4\sqrt{2}h} \left[ \alpha_{(1,\bar{1})}+\alpha_{(0,0)}+\alpha_{(1,0)}+\alpha_{(0,\bar{1})} \right] \left[ \beta_{(1,\bar{1})}-\beta_{(0,0)} \right], \\
&\mathbf{F}_{(\bar{1}/2,\bar{1}/2)} = \frac{1}{4\sqrt{2}h} \left[ \alpha_{(\bar{1},\bar{1})}+\alpha_{(0,0)}+\alpha_{(0,\bar{1})}+\alpha_{(\bar{1},0)} \right] \left[ \beta_{(\bar{1},\bar{1})}-\beta_{(0,0)} \right].
\end{split} \label{flux_11)}
\]

Neither $\widetilde{\mathcal{D}}_{1,0}$ nor $\widetilde{\mathcal{D}}_{0,1}$ is an isotropic discretization of $\widetilde{\mathcal{D}}(\mathbf{r})$. We will show in Section \ref{Fourier_transform} that only their linear combination $\widetilde{\mathcal{D}}_{2,1}$ is isotropic, which utilizes eight fluxes in total, including four in the $\left<10 \right>$ directions and four in the $\left<11 \right>$ directions.

\section{Analyses and results in 2D} \label{Sec:2D_2}
A discretized differential term is isotropic if the leading error does not have preferred directions and anisotropic otherwise \cite{Patra2006, VandeVooren1967, Ananthakrishnaiah1987}. The anisotropy of the leading error can be examined either in the real space or in the Fourier space. In this section, we first briefly review how the anisotropy of a discretization can be examined in the real space, then switch the focus into the Fourier space. Lastly, we implement several finite-difference schemes for PF simulations in 2D.

\subsection{Real space} \label{Sec:lap_real_2D}
The leading error of a discretization usually contains higher-order derivatives and cross derivatives. Those derivatives can be either isotropic or anisotropic. For example, the Laplacian $\nabla^2 \psi=(\psi_{xx}+\psi_{yy})$ is isotropic, but the second derivative $\psi_{xx}$ is anisotropic. This is because $\nabla^2 \psi$ is invariant under coordinate transformation, while $\psi_{xx}$ is not invariant. Thus, we can directly examine the anisotropy of a discretization by rotating the frame of reference. 

For $\mathcal{L}_{1,0}$ in Eq. \eqref{2D_L10}, its Taylor expansion around the origin $(0,0)$ yields
\[
\mathcal{L}_{1,0}=\nabla^2 \psi+\frac{h^2}{12} (\psi_{xxxx}+\psi_{yyyy}) +\mathcal{O}(h^4),
\]
where the second term on the right hand side is the leading error. Because $(\psi_{xxxx}+\psi_{yyyy})$ is not invariant under rotation, $\mathcal{L}_{1,0}$ is an anisotropic discretization. Similarly, the Taylor expansion of $\mathcal{L}_{2,1}$ in Eq. \eqref{2D_L21} gives
\[
\mathcal{L}_{2,1}=\nabla^2 \psi+\frac{h^2}{12} {\nabla}^4 \psi+\mathcal{O}(h^4),
\]
where the second term on the right hand side is the leading error. Since the bilaplacian ${\nabla}^4 \psi$ is invariant under rotation, $\mathcal{L}_{2,1}$ is an isotropic discretization.

In the real space, it is straightforward to examine the anisotropy of a discretization by looking at its leading error. But for differential terms other than Laplacian, the leading errors are nonlinear and usually contain numerous terms, which would yield complex results after rotation. For this reason, we only give the 2D Laplacian as an example in the real space, and will examine the anisotropy of other discretizations in the Fourier space.

\subsection{Fourier space} \label{Fourier_transform}
The leading error of a discretization in the Fourier space usually does not contain spatial derivatives. Thus, the anisotropy can be identified and visualized in a straightforward manner. In this subsection, we discuss discretizations of Laplacian and divergence in the 2D Fourier space.

\subsubsection{Laplacian}
The discrete Fourier transform of the Laplacian is defined as \cite{Thampi2013}
\[
\mathcal{L} (\mathbf{k})=\frac{\sum_{\mathbf{r}} e^{-i\mathbf{k}\cdot\mathbf{r}} \mathcal{L}(\mathbf{r}) } {\sum_{\mathbf{r}} e^{-i\mathbf{k} \cdot\mathbf{r}}\psi(\mathbf{r})}, \label{Fourier_Laplacian}
\]
which has been normalized by the Fourier transform of the scalar field $\psi(\mathbf{r})$, i.e., $\mathcal{F}[\psi(\mathbf{r})]=\sum_{\mathbf{r}} e^{-i\mathbf{k} \cdot\mathbf{r}}\psi(\mathbf{r})$. Since $\mathcal{F}[\psi(\mathbf{r})]$ has no preferred direction in the Fourier space, we can use $\mathcal{L} (\mathbf{k})$ defined in Eq. \eqref{Fourier_Laplacian} to check the anisotropy of a Laplacian in the continuum limit and its finite-difference approximations in the small wave number limit \cite{Thampi2013}. 

\begin{figure}[h]
\centering
\includegraphics[scale=0.65]{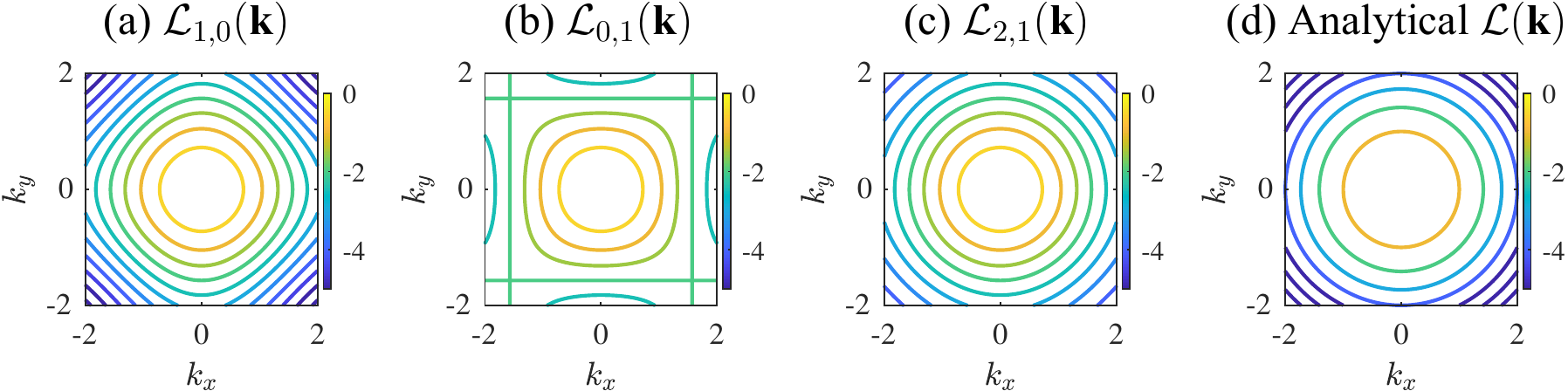}
\caption{Fourier transform of the Laplacian for different discretizations (a)-(c), and for the analytical Laplacian in the continuum limit (d). \label{fig:Fourier_2D_Lap}}
\end{figure}

For a Laplacian in the continuum limit, the Fourier transform gives $\mathcal{L}(\mathbf{k})=-k^2$. For discretizations of Laplacian in Eqs. \eqref{2D_L10}-\eqref{2D_L21}, the Fourier transform in the small wave number limit yields
\[
\mathcal{L}_{1,0} (\mathbf{k})=-k^2+\frac{1}{12}(k^4_x+k^4_y)+\mathcal{O}(k^6_\alpha), \label{L_k_10}
\]
\[
\mathcal{L}_{0,1} (\mathbf{k})=-k^2+\frac{1}{12}(k^4_x+6 k^2_x k^2_y+k^4_y)+\mathcal{O}(k^6_\alpha), \label{L_k_01}
\]
and
\[
\mathcal{L}_{2,1} (\mathbf{k})=-k^2+\frac{1}{12}k^4+\mathcal{O}(k^6_\alpha). \label{L_k_21}
\]
The base discretizations $\mathcal{L}_{1,0}$ and $\mathcal{L}_{0,1}$ are anisotropic because their leading errors at the fourth order of $k$ are not invariant under coordinate transformation. In contrast, $\mathcal{L}_{2,1}$ is isotropic because its leading error $k^4/12=(k^2_x+k^2_y)^2/12$ is invariant. The Fourier transforms of those discretizations are visualized in Fig. \ref{fig:Fourier_2D_Lap}, where $\mathcal{L}_{2,1}$ has improved the isotropy of a discretization in the $k$-space compared with $\mathcal{L}_{1,0}$ and $\mathcal{L}_{0,1}$.

The linear combination $(c_1 \mathcal{L}_{1,0}+c_2 \mathcal{L}_{0,1})$ gives a complete set of discretizations involving both $\left<10\right>$ and $\left<11\right>$ lattice shells. Among them, $\mathcal{L}_{2,1}$ is the only isotropic discretization that can be obtained by solving following equations:
\[
c_1 + c_2 =1, \label{prove1}
\]
and
\[
6 c_2 = 2(c_1+c_2). \label{prove2}
\]
Eq. \eqref{prove1} ensures the sum of weights on two bases equals to 1. Eq. \eqref{prove2} ensures the leading error of the linear combination proportional to $k^4$. The only solutions are $c_1=2/3$ and $c_2=1/3$, which correspond to the isotropic discretization $\mathcal{L}_{2,1}$.

\subsubsection{Divergence} \label{Sec_2D_divergence}
For the generalized divergence $\widetilde{\mathcal{D}}(\mathbf{r})=\vec{\nabla} \cdot ( \alpha {\vec{\nabla}\beta} )$, the discrete Fourier transform is defined as
\[
\widetilde{\mathcal{D}} (\mathbf{k})=\frac{\sum_{\mathbf{r}} e^{-i\mathbf{k}\cdot\mathbf{r}} \widetilde{\mathcal{D}} (\mathbf{r}) } {\sum_{\mathbf{r}} e^{-i\mathbf{k} \cdot\mathbf{r}}\alpha(\mathbf{r}) \beta(\mathbf{r})}. \label{Fourier_divergence}
\]
It is normalized by the convolution of two functions in the Fourier space, i.e., $\mathcal{F}[\alpha(\mathbf{r})\beta(\mathbf{r})]=\sum_{\mathbf{k}^{\prime}} \left[ A(\mathbf{k}^{\prime}) \cdot B(\mathbf{k}-\mathbf{k}^{\prime}) \right] =A(\mathbf{k})*B(\mathbf{k})$, where $A(\mathbf{k})=\mathcal{F}[\alpha(\mathbf{r})]=\sum_{\mathbf{r}} e^{-i\mathbf{k} \cdot\mathbf{r}}\alpha(\mathbf{r})$ and $B(\mathbf{k})=\mathcal{F}[\beta(\mathbf{r})]=\sum_{\mathbf{r}} e^{-i\mathbf{k} \cdot\mathbf{r}}\beta(\mathbf{r})$. Since the Fourier transform of a scalar field $A(\mathbf{k})$ or $B(\mathbf{k})$ has no preferred direction in the Fourier space, we can use $\widetilde{\mathcal{D}} (\mathbf{k})$ defined in Eq. \eqref{Fourier_divergence} to check the anisotropy of a divergence in the continuum limit and its finite-difference approximations in the small wave number limit.

\begin{figure}[h]
\centering
\includegraphics[scale=0.65]{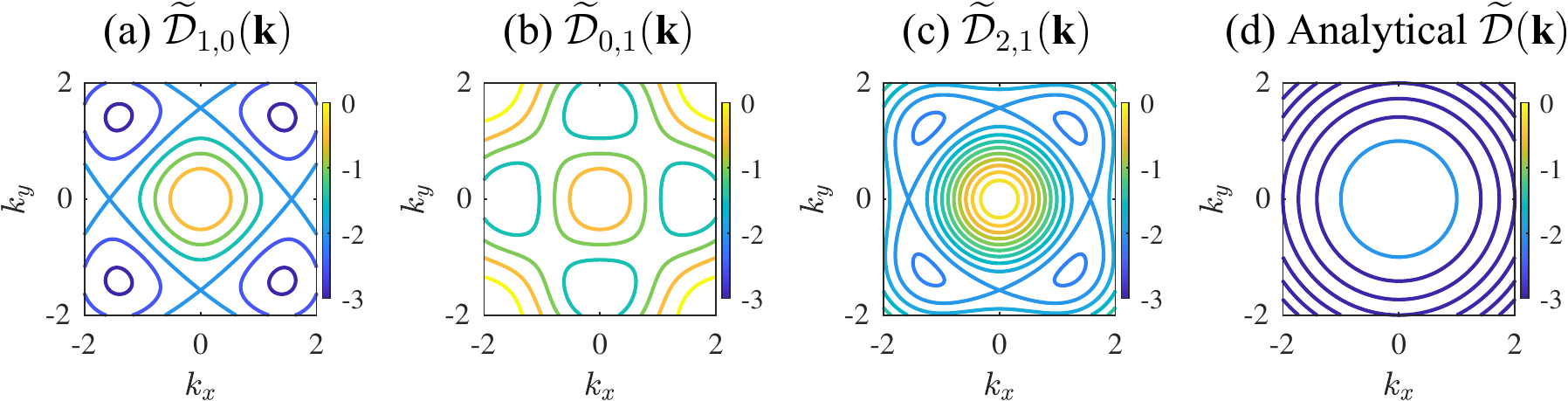}
\caption{Fourier transform of the divergence for different discretizations (a)-(c), and for the analytical divergence in the continuum limit (d). \label{fig:Fourier_2D_Div}}
\end{figure}

For a divergence in the continuum limit, the Fourier transform gives $\widetilde{\mathcal{D}}(\mathbf{k})=-2k^2$. For discretizations of the divergence, the Fourier transform in the small wave number limit gives
\[
\widetilde{\mathcal{D}}_{1,0}(\mathbf{k})=-2k^2+\frac{2}{3}\left(k_x^4+\frac{3}{4}k_x^2k_y^2+k_y^4 \right) +\mathcal{O}(k^6_\alpha), \label{D_k_10}
\]
\[
\widetilde{\mathcal{D}}_{0,1}(\mathbf{k})=-2k^2+\frac{2}{3}\left(k_x^4+\frac{9}{2}k_x^2k_y^2+k_y^4 \right) +\mathcal{O}(k^6_\alpha), \label{D_k_01}
\]
and
\[
\widetilde{\mathcal{D}}_{2,1}(\mathbf{k})=-2k^2+\frac{2}{3}k^4 +\mathcal{O}(k^6_\alpha). \label{D_k_21}
\]
The base discretizations $\widetilde{\mathcal{D}}_{1,0}$ and $\widetilde{\mathcal{D}}_{0,1}$ are anisotropic because their leading errors at the fourth order of $k$ are not invariant under coordinate transformation. $\widetilde{\mathcal{D}}_{2,1}$ is isotropic because its leading error $2 k^4/3$ is invariant. The Fourier transforms of those discretizations are visualized in Fig. \ref{fig:Fourier_2D_Div}. Similar to the Laplacian, one can show that $\widetilde{\mathcal{D}}_{2,1}$ is the only isotropic discretization of the divergence through linear combination of $\widetilde{\mathcal{D}}_{1,0}$ and $\widetilde{\mathcal{D}}_{0,1}$. Although the coefficients of the leading errors at order $k^4$ in Eqs. \eqref{D_k_10}-\eqref{D_k_01} are different from that in Eqs. \eqref{L_k_10}-\eqref{L_k_01}, the weights on two bases for an isotropic discretization are the same, i.e., $2/3$ weight on the $\left<10\right>$ base and $1/3$ weight on the $\left<11\right>$ base.

\subsection{Phase-field simulations in 2D} \label{Sec:2D_results}

\subsubsection{Numerical implementation} \label{Sec:2D_process}
In this section, we aim at characterizing effects of finite-difference schemes on PF simulations of crystal growth. A finite-difference scheme is the collective discretization of leading differential terms in the equations of motion, including $\mathcal{L}(\mathbf{r})$, $\mathcal{G}(\mathbf{r})$, $\mathcal{D}(\mathbf{r})$, and the anti-trapping. We use $\mathcal{S}_{s_1,s_2}$ to denote a finite-difference scheme with the standard anti-trapping $\mathcal{A}(\mathbf{r})$ in Eq. \eqref{Anti_convention}, and use $\bar{\mathcal{S}}_{s_1,s_2}$ to denote a finite-difference scheme with the approximated anti-trapping $\bar{\mathcal{A}}(\mathbf{r})$ in Eq. \eqref{Anti_approx}. The subscripts $s_1$ and $s_2$ indicate the ratio of bases for each leading differential terms as in Eq. \eqref{tau_2D}. Note that we use the method of Ref. \cite{Tourret2015} to discretize divergences of the supersaturation gradient and the anti-trapping current for the anisotropic scheme ${\mathcal{S}}_{1,0}$, where only the $\left<10\right>$ lattice shell is involved. For $\bar{\mathcal{A}}_{2,1}$ in the isotropic scheme $\bar{\mathcal{S}}_{2,1}$, it has a similar form as $\widetilde{\mathcal{D}}_{2,1}$; but for ${\mathcal{A}}_{2,1}$ in the isotropic scheme ${\mathcal{S}}_{2,1}$, it includes the discretization of the gradient that has vanishing second-order error as discussed in Appendix \ref{Sec:appendix_vanish}. Meanwhile, for discretizations of other differential terms at the first or higher orders of $\epsilon_4$ in the evolution equation of the PF, only the $\left<10\right>$ lattice shell is utilized as in Ref. \cite{Tourret2015}.

We implement three finite-difference schemes ${\mathcal{S}}_{1,0}$, ${\mathcal{S}}_{2,1}$, and $\bar{\mathcal{S}}_{2,1}$ for PF simulations of both isothermal crystal growth and directional solidification. The Euler explicit method is used for discretization in time, where a time step $\Delta t=0.8 h^2/4D$ is chosen to ensure numerical stability. Massively parallel simulations were performed on GPU with the computer unified device architecture (CUDA) programming language. The time loop is comprised of two main CUDA kernels for updating $\psi$ and $U$ fields, and the time stepping is achieved by swapping pointer addresses of arrays that contain each $\psi$ and $U$ fields at the current and the next time steps.

\subsubsection{Isothermal solidification}
\begin{figure}[h]
\centering
\includegraphics[scale=0.3]{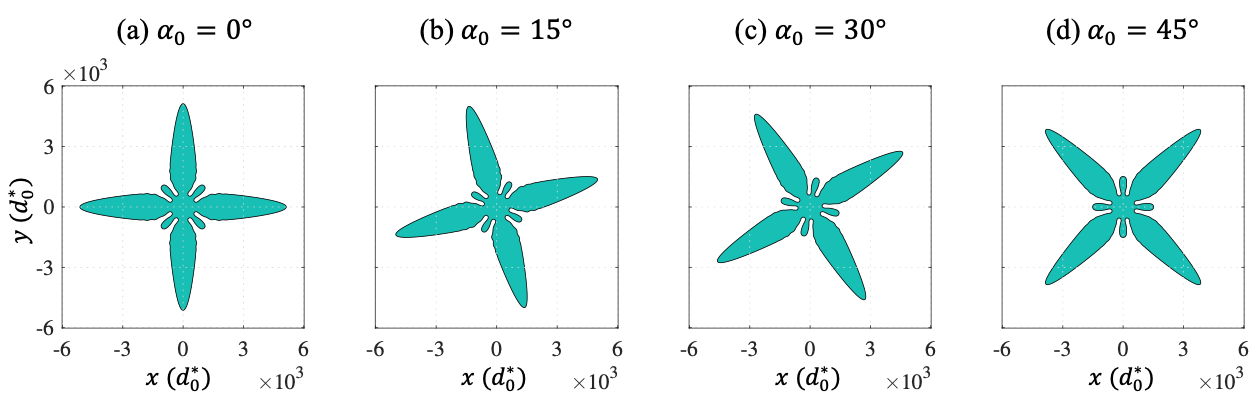}
\caption{The isothermal crystal growth in a supersaturated melt with different orientations. \label{fig:2D_Isothermal_Example}}
\end{figure}

For the isothermal crystal growth, a circular seed of radius $r=50$ $(d_0^*)$ is initially placed at the center of a supersaturated melt with the scaled supersaturation $\Omega=0.43$, where the capillary length $d_0^*$ and $\Omega$ are defined in \ref{Sec:evolution_eqs}. The no-flux conditions are applied at boundaries of a square simulation domain with side length $1.5\times 10^4$ $(d_0^*)$. The total simulation time is set to $t=1.5 \times 10^6$ $({d_0^*}^2/D)$, which ensures that the diffusion field does not interact with boundaries. A series of simulations were carried out to explore the influence of $\alpha_0$, the angle between the preferred crystal growth direction and the $y$-axis. It varies between $0^{\circ}$ and $45^{\circ}$ due to the four-fold symmetry of the anisotropy. Examples of isothermal dendrite growth with different $\alpha_0$ are given in Fig. \ref{fig:2D_Isothermal_Example}, where the isotropic scheme $\bar{\mathcal{S}}_{2,1}$ is used with $W/d_0^*=16$, $h/W=1.2$, and $\epsilon_4=0.014$.

According to the solvability theory, the scaling parameter $\sigma^*=2 D d_0^*/\rho^2V$ (where $\rho$ is the tip radius and $V$ is the tip growth velocity) approaches a constant at low undercooling that only depends on the anisotropy strength \cite{Barbieri1989,Karma2000}, which is also known as the tip selection constant. In order to measure $\sigma^*$, we track the position of the most advanced solid-liquid interface along the $y$-direction, and acquire its location ($x_{tip}$, $y_{tip}$). We also track the tip radius by following three steps: firstly, use a MATLAB script to read the final output of $\psi$-field and find the contour of $\psi=0$; then rotate the contour by $(-\alpha)$ such that the four arms of the dendrite are aligned to the grid, where $\alpha$ is the actually crystal growth angle measured by tracking the tip position; lastly, we extract the tip radius $\rho$ by fitting the interface shape to a parabola $y=y_{tip}-(x-x_{tip})^2/(2\rho)$ as described in Ref. \cite{Clarke2017}.

\begin{figure}[h]
\centering
\includegraphics[scale=0.3]{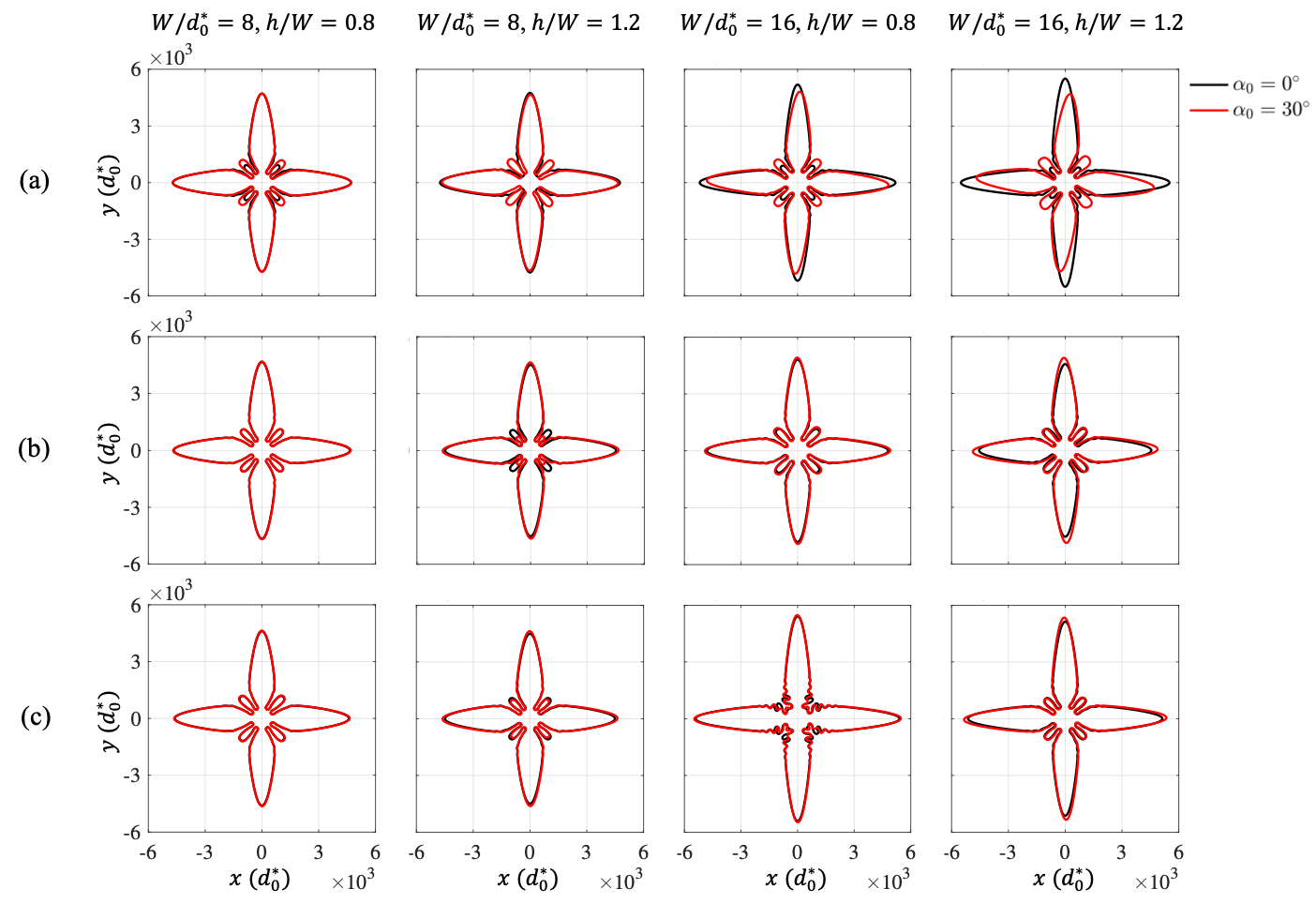}
\caption{Comparison of 2D PF simulations with different crystal orientations $\alpha_0=0^{\circ}$ and $\alpha_0=30^{\circ}$. Simulations use three finite-difference schemes: (a) the anisotropic scheme $\mathcal{S}_{1,0}$, (b) the isotropic scheme $\mathcal{S}_{2,1}$, and (c) the isotropic scheme $\bar{\mathcal{S}}_{2,1}$. The contours of $\alpha_0=30^{\circ}$ are rotated back by $(-\alpha_0)$ to be compared with the contours of $\alpha_0=0^{\circ}$. \label{fig:2D_examples}}
\end{figure}

PF simulations with different crystal orientations $\alpha_0=0^{\circ}$ and $\alpha_0=30^{\circ}$ are compared in Fig. \ref{fig:2D_examples}. It shows a direct evidence that isotropic finite-difference schemes can improve the rotational invariance of PF simulations. Note that the crystal growth is symmetric to the lattice grid with $\alpha_0=0^{\circ}$, while asymmetric with $\alpha_0=30^{\circ}$. In principle, the crystal morphology should be invariant under rotation, i.e., the superposed solid-liquid interfaces of $\alpha_0=0^{\circ}$ and $\alpha_0=30^{\circ}$ should be the same. However, the anisotropic scheme $\mathcal{S}_{1,0}$ breaks the rotational invariance as shown in Fig. \ref{fig:2D_examples}(a), especially for increased interface thickness $W$ and lattice spacing $h$. Meanwhile, those two isotropic schemes $\mathcal{S}_{2,1}$ and $\bar{\mathcal{S}}_{2,1}$ manage to better maintain the rotational invariance for increased $W$ and $h$, as shown in Fig. \ref{fig:2D_examples}(b)-(c).

\begin{figure}[h]
\centering
\includegraphics[scale=0.65]{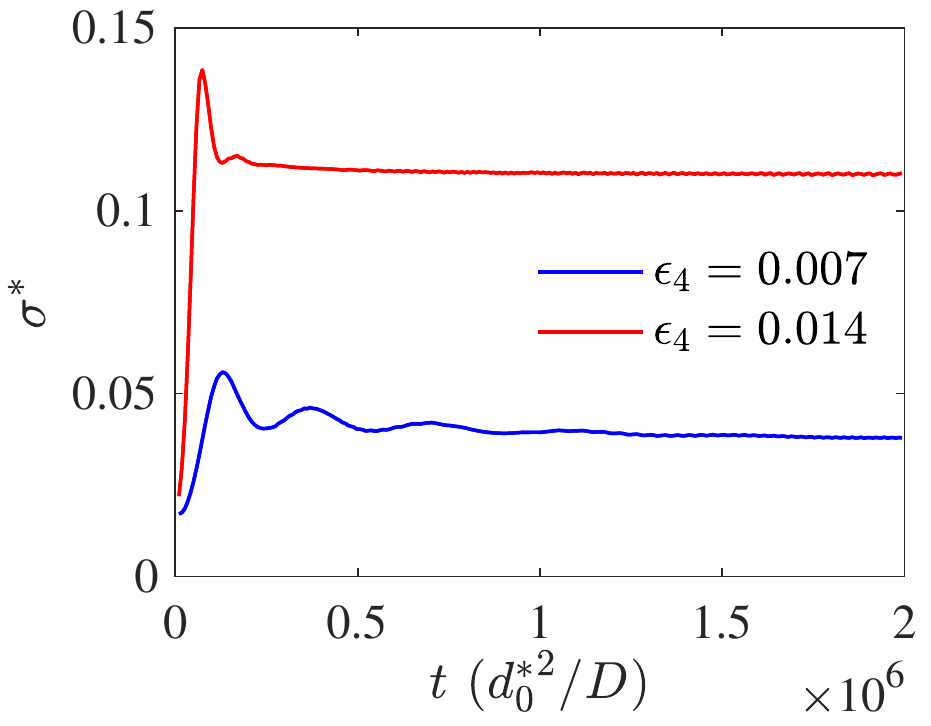}
\caption{The tip selection constant $\sigma^*$ as a function of time in 2D PF simulations using the $\mathcal{S}_{1,0}$ scheme, with $\alpha_0=0^{\circ}$, $W/d_0^*=8$, and $h/W= 0.8$. \label{fig:2D_sigma_t}}
\end{figure}

\begin{figure}[h]
\centering
\includegraphics[scale=0.65]{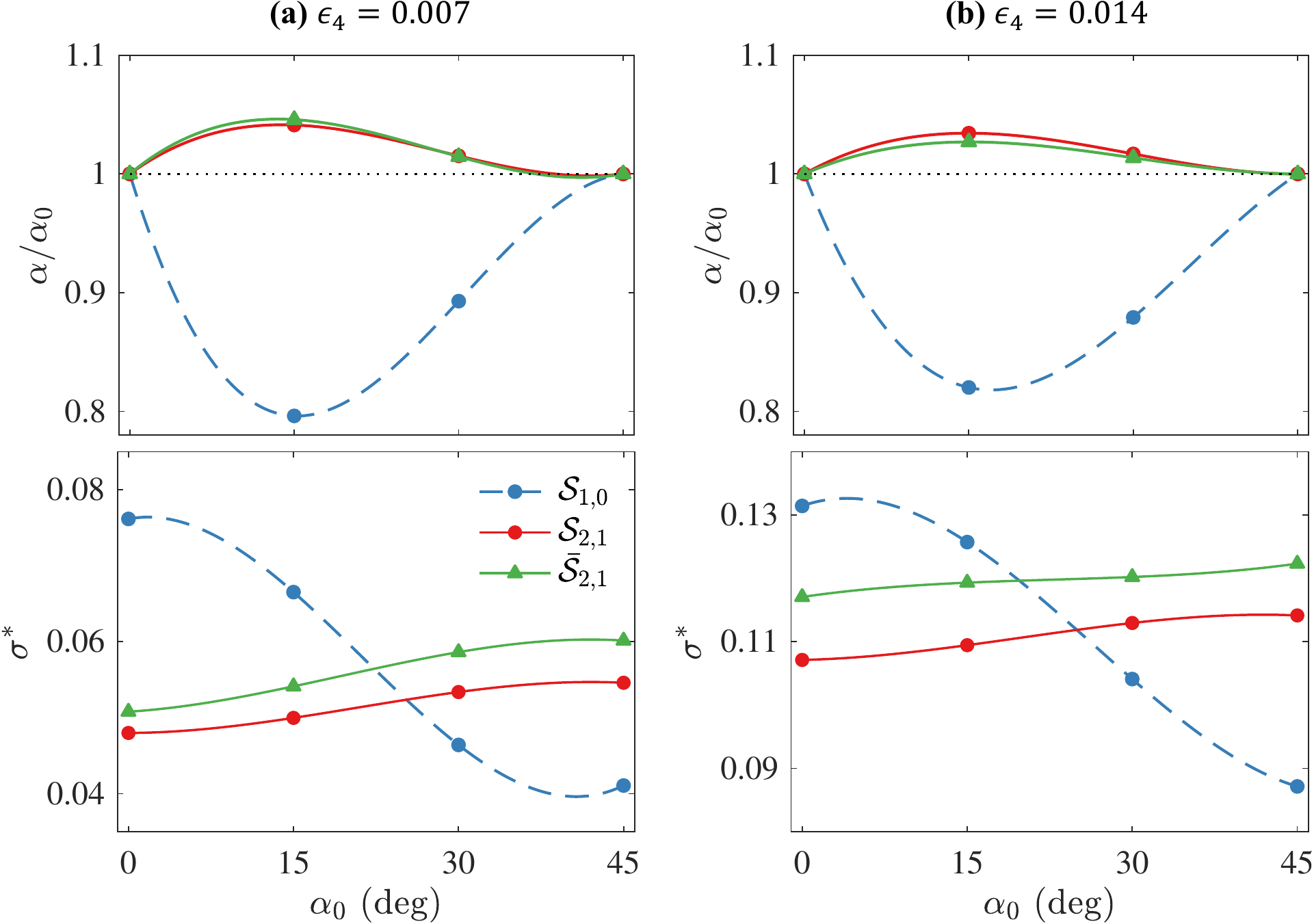}
\caption{Measurements of $\alpha/\alpha_0$ and $\sigma^*$ as a function of $\alpha_0$ for PF simulations of isothermal crystal growth. The isotropic schemes $\mathcal{S}_{2,1}$ and $\bar{\mathcal{S}}_{2,1}$ can significantly improve the rotational invariance in terms of both $\alpha/\alpha_0$ and $\sigma^*$. \label{fig:Plot_2D_angles}}
\end{figure}

Effects of finite-difference schemes are further examined by two criteria, the scaled crystal growth direction $\alpha/\alpha_0$ and the tip selection constant $\sigma^*$. The value of $\alpha/\alpha_0$ provides a way to know the strength of the lattice anisotropy, which tends to push $\alpha$ away from $\alpha_0$. The value of $\sigma^*$ in principle is invariant under rotation, which becomes almost constant long before the tip velocity and radius have reached their steady-state values \cite{Plapp2000}. The time-dependence of $\sigma^*$ in 2D PF simulations with the scheme $\mathcal{S}_{1,0}$ and $\alpha_0=0^{\circ}$ is shown in Fig. \ref{fig:2D_sigma_t}, where we measured $\sigma^*$ at $t=1.5 \times 10^6$ $({d_0^*}^2/D)$. We measured $\sigma^*$ at the same time when it reached steady-state in the other PF simulations with different schemes and crystal orientations. Fig. \ref{fig:Plot_2D_angles} shows $\alpha/\alpha_0$ and $\sigma^*$ as a function of $\alpha_0$, where PF simulations use parameters $W/d_0^*=16$ and $h/W=1.2$. The isotropic schemes $\mathcal{S}_{2,1}$ and $\bar{\mathcal{S}}_{2,1}$ yielded similar results and have significantly improved the rotational invariance in terms of both criteria, which also indicate that PF simulations with the approximated anti-trapping current are quantitatively accurate.

The performance of finite-difference schemes is summarized in Table \ref{tab:2D_performance}, where we measured the efficiency of PF simulations in Fig. \ref{fig:Plot_2D_angles}. Because two isotropic schemes $\mathcal{S}_{2,1}$ and $\bar{\mathcal{S}}_{2,1}$ require more computation, they are less efficient than the anisotropic scheme $\mathcal{S}_{1,0}$. Among them, $\bar{\mathcal{S}}_{2,1}$ is slightly more efficient than $\mathcal{S}_{2,1}$ since the discretization of the approximated anti-trapping requires less computation.

\begin{table}[h]
\caption{The performance of three finite-different schemes in 2D. The efficiency measures the time a PF simulation takes on an NVIDIA Tesla P100 GPU. \label{tab:2D_performance}}
\centering
\begin{tabular}{cccc} \hline
Scheme & Anti-trapping & Property & Efficiency ($s$)\\\hline
$\mathcal{S}_{1,0}$ & ${\mathcal{A}}(\mathbf{r})$ & Anisotropic & 1359 \\
$\mathcal{S}_{2,1}$ & ${\mathcal{A}}(\mathbf{r})$ & Isotropic & 4612 \\
$\bar{\mathcal{S}}_{2,1}$ & $\bar{\mathcal{A}}(\mathbf{r})$ & Isotropic & 3883 \\
\hline
\end{tabular}
\end{table}


\subsubsection{Directional solidification}
\label{sec:directional_2D}

For directional solidification, the crystal grows within a temperature gradient $G$ at an imposed pulling velocity $V_p$. Here the crystalline misorientation $\alpha_0$ is defined as the angle between the $\left<100\right>$ preferred growth direction and $G$. Due to the effect of $G$, the crystal growth angle $\alpha/\alpha_0$ is generally a function of the local primary spacing \cite{Deschamps2008,Ghmadh2014,Song2018}. But in the well-developed dendritic regime we consider here, the growth of dendritic tips closely follows the crystalline orientation and $\alpha \approx \alpha_0$.

The growth competition of columnar dendritic grains with different $\alpha_0$ determines the formation and evolution of grain boundaries, which have a significant influence on the mechanical behavior of metallic materials formed by directional solidification. The grain competition at a relatively small $V_p$ was studied by 2D PF simulations with the conventional finite-difference scheme $\mathcal{S}_{1,0}$ \cite{Tourret2015,Pineau2018}. Although $\mathcal{S}_{1,0}$ is generally sufficient for those studies, it becomes inadequate for quantitatively modeling grain competition in the well-developed dendritic regime at a relatively large $V_p$ \cite{Dorari2021}. In order to demonstrate the effects of finite-difference schemes in the latter scenario, we simply consider the PF simulations of directional solidification of a single crystal with different $\alpha_0$. According to those simulations, we can obtain a characteristic length scale $\delta$, defined as the distance between the most advanced tips of the well-oriented grain ($\alpha_0=0^{\circ}$) and the misoriented grain ($\alpha_0>0^{\circ}$), using the relation
\[
\delta=\left(\Delta_1-\Delta_0 \right) \cdot l_T,
\]
where $\Delta_1$ and $\Delta_0$ are dimensionless tip undercooling $\Delta=(T_l-T)/\Delta T_0$ for the misoriented and well-oriented grains, respectively, and $T_l$ is the liquidus temperature at $c=c_{\infty}$. Then, the measured $\delta$ in PF simulations can be compared with the analytical prediction from the Ivantsov-solvability theory as introduced in Ref. \cite{Dorari2021}.

\begin{figure}[h]
\centering
\includegraphics[scale=0.3]{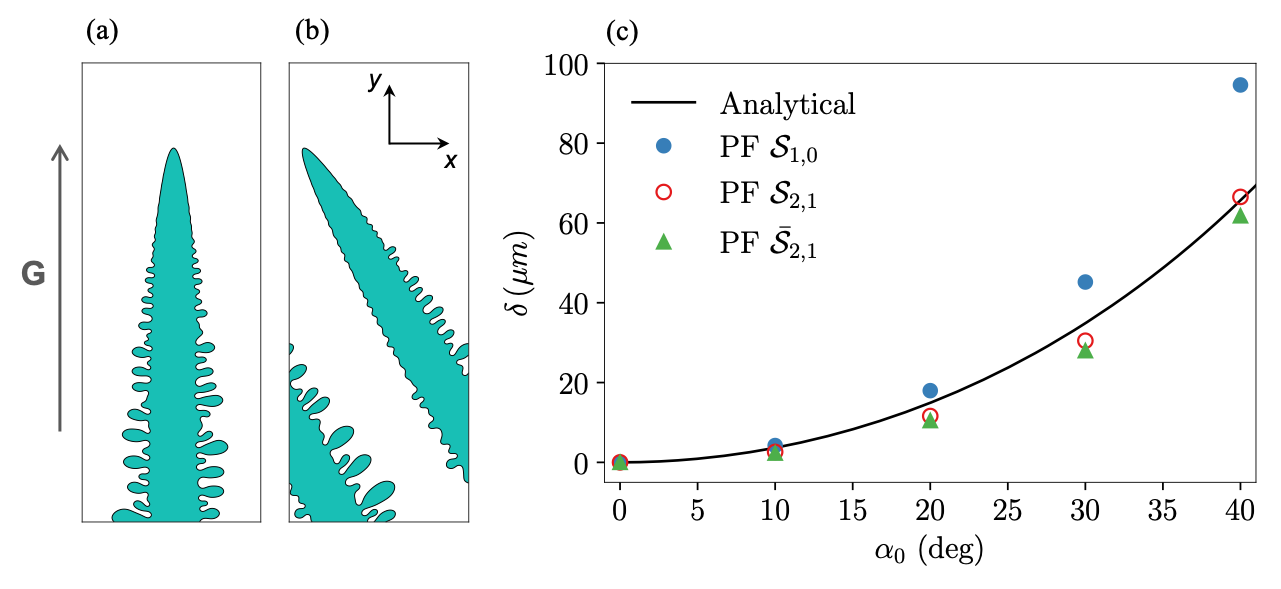}
\caption{2D PF simulations of directional solidification. The isotropic finite-difference scheme $\mathcal{S}_{2,1}$ is used for simulations with a misorientation angle $\alpha_0$ between the preferred $\left<100\right>$ growth direction and the temperature gradient: (a) $\alpha_0=0^{\circ}$ and (b) $\alpha_0=30^{\circ}$. The simulation domain size is $L_x\times L_y = 191\times491$ $\mathrm{\mu m^2}$ with periodic boundary conditions applied in the $x$ directions. (c) The characteristic length scale $\delta$ as a function of $\alpha_0$. Results of PF simulations using different schemes (symbols) are compared with the analytical prediction of the Ivantsov-solvability theory (line). The analytical solution considers a constant $\sigma=0.0973$ that is derived analytically for a one-sided solidification model with $\epsilon_4=0.014$ \cite{Barbieri1989}. \label{fig:2D_DS}}
\end{figure}

Results of 2D PF simulations are shown in Fig. \ref{fig:2D_DS}. Those simulations use a relatively high pulling velocity $V_p=86$ $\mathrm{\mu m/s}$ with $G=19$ $\mathrm{K/cm}$ for directional solidification of a SCN-1.3 wt.\% acetone alloy. The other simulation parameters used in Fig. \ref{fig:2D_DS} include: the interface thickness $W/d_0 = 110.4$, the grid spacing $h/W = 1.2$, the explicit time step $t=1 \times 10^{-5}$ s, the crystalline anisotropy $\epsilon_4=0.014$, the partition coefficient $k=0.1$, the liquidus slope $m = -3.02$ $\mathrm{K/wt.\%}$, and the diffusion coefficient $D = 1270$ $\mathrm{\mu m^2/s}$. The simulations use a domain of $L_x\times L_y = 191\times491$ $\mathrm{\mu m^2}$ with periodic boundary conditions applied in the $x$ directions. Thus, the primary spacing of the dendrite is 191 $\mathrm{\mu m}$, which is large enough compared with the diffusion length $l_D=D/V_p=15$ $\mathrm{\mu m}$, so that the tip of a dendrite is not affected by its effective neighbors due to the periodic boundary conditions. 

As shown in Fig. \ref{fig:2D_DS}(c), the isotropic schemes $\mathcal{S}_{2,1}$ and $\bar{\mathcal{S}}_{2,1}$ yield similar results that have better agreement with the analytical solution than the anisotropic scheme $\mathcal{S}_{1,0}$, especially for large misorientation angles. For quantitative modeling of grain competition in the well-developed dendritic regime, accurately resolving the characteristic length scale $\delta$ is critical as it can influence the grain-boundary dynamics, for which using the isotropic finite-difference implementation is essentially necessary. A detailed study of grain competition using the isotropic scheme $\mathcal{S}_{2,1}$ will be presented elsewhere \cite{Dorari2021}.

\section{Isotropic discretizations in 3D} \label{Sec:3D_1}
\begin{figure}[h]
\centering
\includegraphics[scale=0.65]{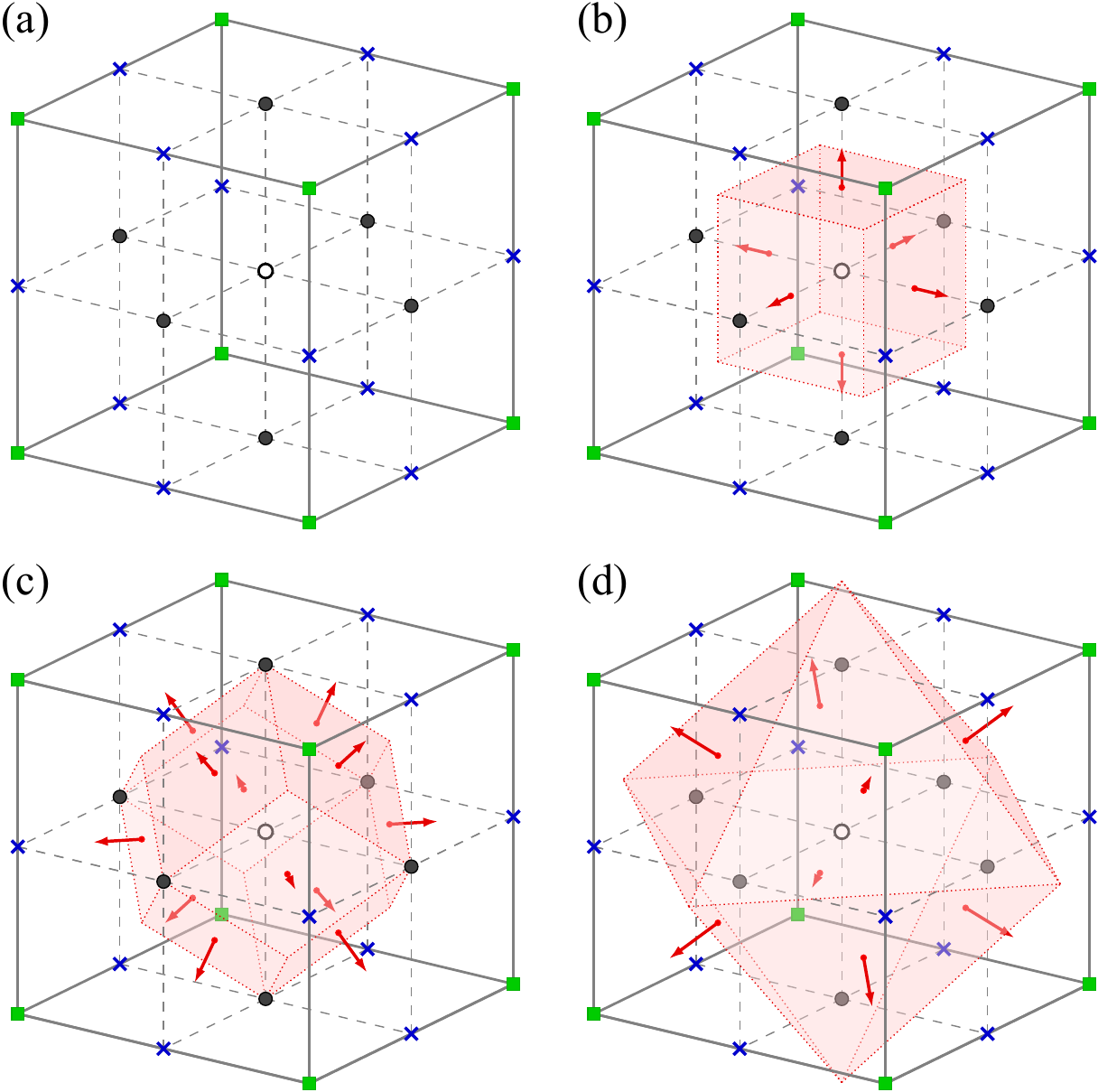}
\caption{(a) A unit cell in the 3D cubic lattice. The origin point is marked by a black hallow circle. The $\left<100\right>$ lattice shell is marked by black filled circles, the $\left<110\right>$ lattice shell is marked by blue crosses, and the $\left<111\right>$ lattice shell is marked by green squares. The fluxes in the (b) $\left<100\right>$ directions, (c) $\left<110\right>$ directions, and (d) $\left<111\right>$ directions for evaluating a divergence.
\label{fig:3D_Fluxes}}
\end{figure}

We first introduce a notation system that can be used to describe the discretization of all leading differential terms in 3D. Consider an arbitrary differential term $\mathcal{T}(\mathbf{r})$ in 3D. It is discretized by finite difference on a cubic lattice with the lattice spacing $\Delta x=\Delta y=\Delta z=h$, as shown in Fig. \ref{fig:3D_Fluxes}(a). The coordinates of a lattice point $(r_1 h,r_2 h,r_3 h)$ are abbreviated as $(r_1,r_2,r_3)$. Similar to notations in 2D, we use $\mathcal{T}_{1,0,0}(\mathbf{r})$ to denote the discretization involving only the $\left<100\right>$ lattice shell, $\mathcal{T}_{0,1,0}(\mathbf{r})$ to denote the discretization involving only the $\left<110\right>$ lattice shell, and $\mathcal{T}_{0,0,1}(\mathbf{r})$ to denote the discretization involving only the $\left<111\right>$ lattice shell. Note that the origin point is involved by default in all cases. Since each lattice shell can be used independently to discretize $\mathcal{T}(\mathbf{r})$, we call $\mathcal{T}_{1,0,0}(\mathbf{r})$, $\mathcal{T}_{0,1,0}(\mathbf{r})$, and $\mathcal{T}_{0,0,1}(\mathbf{r})$ base discretizations. These three bases can be combined to produce a complete set of discretizations involving $\left<100\right>$, $\left<110\right>$, and $\left<111\right>$ lattice shells. Thus, we can use $\mathcal{T}_{s_1,s_2,s_3}(\mathbf{r})$, or simply $\mathcal{T}_{s_1,s_2,s_3}$ to denote an arbitrary discretization with subscript indicating the ratio of three bases, such that
\[
\mathcal{T}_{s_1,s_2,s_3}=\frac{s_1}{s_1+s_2+s_3} \mathcal{T}_{1,0,0} + \frac{s_2}{s_1+s_2+s_3} \mathcal{T}_{0,1,0}+ \frac{s_3}{s_1+s_2+s_3} \mathcal{T}_{0,0,1}. \label{tau_3D}
\]
A bar on the subscript is used to denote negative, for example $\bar{s}_1$ represents $(-s_1)$. Note that the sum of weights on three bases always equals to 1. In the following, the same notations are applied to all differential terms in 3D.

\subsection{Laplacian}
The most commonly used discrete Laplacian employs only the $\left<100\right>$ lattice shell. It is referred to as $\mathcal{L}_{1,0,0}$ in this context, and has the form
\[
\mathcal{L}_{1,0,0}=\frac{1}{h^2} \left[ \sum_{i=1}^6 \psi_{\left<100\right>}^i-6 \psi_{\left(0,0,0\right)} \right], \label{3D_L100}
\]
where the index $i$ sums over values of $\psi$ in the $\left<100\right>$ lattice shell, i.e., $\psi_{\left<100\right>}^i$. Here $\mathcal{L}_{1,0,0}$ constitutes one base for discretizing $\mathcal{L}(\mathbf{r})$. The other two bases employ $\left<110\right>$ and $\left<111\right>$ lattice shells, respectively, and have forms
\[
\mathcal{L}_{0,1,0}=\frac{1}{4h^2} \left[\sum_{i=1}^{12} \psi_{\left<110\right>}^i-12 \psi_{\left(0,0,0\right)} \right], \label{3D_L110}
\]
and
\[
\mathcal{L}_{0,0,1}=\frac{1}{4h^2} \left[\sum_{i=1}^{8} \psi_{\left<111\right>}^i-8 \psi_{\left(0,0,0\right)} \right]. \label{3D_L111}
\]
The above three base discretizations of the Laplacian are anisotropic. As we will discuss in Section \ref{Sec:3D_Lap_Fourier}, there are various combinations of those three bases to obtain an isotropic Laplacian at order $h^2$. Among them, only one combination utilizes the first two bases with lattice shells in the $\left<100\right>$ and $\left<110\right>$ directions, which is $\mathcal{L}_{1,2,0}$ following the notation defined in Eq. \eqref{tau_3D}, or referred to as the 18-point formula in Ref. \cite{Glasner2001}.

\subsection{Divergence} \label{Sec:3D_Div_dis}
For the generalized divergence $\widetilde{\mathcal{D}}(\mathbf{r})=\vec{\nabla} \cdot \mathbf{F}=\vec{\nabla}\cdot( \alpha {\vec{\nabla}\beta} )$ in 3D, there are three base discretizations with different flux orientations. The first base utilizes six fluxes of $\mathbf{F}$ in $\langle 100 \rangle$ directions, as shown in Fig. \ref{fig:3D_Fluxes}(b). The region of space (shaded region) for evaluating the $\left<100\right>$ fluxes is a cube with an edge length $h$. This base is denoted by $\widetilde{\mathcal{D}}_{1,0,0}$, and has the form  
\[
\widetilde{\mathcal{D}}_{1,0,0} = \frac{1}{h} \left [ \mathbf{F}_{(1/2,0,0)}+\mathbf{F}_{(\bar{1}/2,0,0)}+\mathbf{F}_{(0,1/2,0)}+\mathbf{F}_{(0,\bar{1}/2),0}+\mathbf{F}_{(0,0,1/2)}+\mathbf{F}_{(0,0,\bar{1}/2)}  \right], \label{divergence_100}
\]
where $\mathbf{F}_{(i,j,k)}$ is the flux at the off-lattice position $(i,j,k)$, and the flux pointing out of the unit cell is positive. To evaluate each flux, a properly averaged $\alpha$ at the off-lattice location $(i,j,k)$, i.e., $\bar{\alpha}_{(i,j,k)}$, is necessary. This average takes $\alpha$ values at the nearest and the next-nearest lattice points around the position $(i,j,k)$. In order to obtain an isotropic discretization of $\widetilde{\mathcal{D}}(\mathbf{r})$ later, we propose to evaluate one of the $\left<100\right>$ fluxes as
\[
\mathbf{F}_{(1/2,0,0)}=\bar{\alpha}_{(1/2,0,0)}\frac{\left[\beta_{(1,0,0)}-\beta_{(0,0,0)}\right]}{h},
\]
where the value of $\bar{\alpha}_{(1/2,0,0)}$ is taken by putting $1/4$ weight on the two nearest lattice points and $1/16$ weight on the eight next-nearest lattice points, i.e.,
\[
\begin{split}
\bar{\alpha}_{(1/2,0,0)}=&\frac{1}{4}\left[\alpha_{(1,0,0)}+\alpha_{(0,0,0)}\right]
+\frac{1}{16}\left[\alpha_{(1,1,0)}+\alpha_{(0,1,0)}+\alpha_{(0,0,1)}\right.\\
&\left.+\alpha_{(1,0,1)}+\alpha_{(1,\bar{1},0)}+\alpha_{(0,\bar{1},0)}+\alpha_{(0,0,\bar{1})}+\alpha_{(1,0,\bar{1})}\right].
\end{split}
\]
Similarly, the other five $\left<100\right>$ fluxes in Eq. \eqref{divergence_100} are evaluated in this manner.

The second base discretization of $\widetilde{\mathcal{D}}(\mathbf{r})$ utilizes twelve fluxes of $\mathbf{F}$ in the $\langle 110 \rangle$ directions, as shown in Fig. \ref{fig:3D_Fluxes}(c). The region of space (shaded region) for evaluating the $\left<110\right>$ fluxes is a rhombic dodecahedron with an edge length $(\sqrt{3}h/2)$. For such an irregular region of space, we use a coordinate invariant form of the discrete divergence \cite{Margolin2000}. The second base discretization is denoted by $\widetilde{\mathcal{D}}_{0,1,0}$, and has the form
\[
\begin{split}
\widetilde{\mathcal{D}}_{0,1,0} = \frac{1}{2\sqrt{2}h} \big[ &\mathbf{F}_{(1/2,1/2,0)}+\mathbf{F}_{(1/2,\bar{1}/2,0)}+\mathbf{F}_{(\bar{1}/2,1/2,0)}+\mathbf{F}_{(\bar{1}/2,\bar{1}/2,0)} \\
+&\mathbf{F}_{(1/2,0,1/2)}+\mathbf{F}_{(1/2,0,\bar{1}/2)}+\mathbf{F}_{(\bar{1}/2,0,1/2)}+\mathbf{F}_{(\bar{1}/2,0,\bar{1}/2)}\\
+&\mathbf{F}_{(0,1/2,1/2)}+\mathbf{F}_{(0,1/2,\bar{1}/2)}+\mathbf{F}_{(0,\bar{1}/2,1/2)}+\mathbf{F}_{(0,\bar{1}/2,\bar{1}/2)} \big]. \label{divergence_110}
\end{split}
\]
One of the $\langle 110 \rangle$ fluxes is evaluated by
\[
\mathbf{F}_{(1/2,1/2,0)}=\bar{\alpha}_{(1/2,1/2,0)}\frac{\left[\beta_{(1,1,0)}-\beta_{(0,0,0)}\right]}{\sqrt{2}h},
\]
where the value of $\bar{\alpha}_{(1/2,1/2,0)}$ is taken by putting $3/16$ weight on the four nearest lattice points and $1/32$ weight on the eight next-nearest lattice points, i.e.,
\[
\begin{split}
\bar{\alpha}_{(1/2,1/2,0)}=&\frac{3}{16}\left[\alpha_{(1,0,0)}+\alpha_{(1,1,0)}+\alpha_{(0,1,0)}+\alpha_{(0,0,0)}\right]+\frac{1}{32}\left[\alpha_{(1,1,1)}+\alpha_{(1,0,1)}\right. \\
&\left.+\alpha_{(0,1,1)}+\alpha_{(0,0,1)}+\alpha_{(1,1,\bar{1})}+\alpha_{(1,0,\bar{1})}+\alpha_{(0,1,\bar{1})}+\alpha_{(0,0,\bar{1})}\right].
\end{split}
\]
Similarly, the other eleven $\langle 110 \rangle$ fluxes in Eq. \eqref{divergence_110} are evaluated in this manner.

The third base discretization of $\widetilde{\mathcal{D}}(\mathbf{r})$ utilizes eight fluxes of $\mathbf{F}$ in the $\langle 111 \rangle$ directions, as shown in Fig. \ref{fig:3D_Fluxes}(d). The region of space (shaded region) for evaluating the $\left<111\right>$ fluxes is a regular octahedron with an edge length $(3\sqrt{2}h/2)$. We use the same coordinate-invariant form of the discrete divergence \cite{Margolin2000}. The third base discretization is denoted by $\widetilde{\mathcal{D}}_{0,0,1}$, and has the form
\[
\begin{split}
\widetilde{\mathcal{D}}_{0,0,1} = \frac{\sqrt{3}}{4h} \big[ &\mathbf{F}_{(1/2,1/2,1/2)}+\mathbf{F}_{(\bar{1}/2,1/2,1/2)}+\mathbf{F}_{(1/2,\bar{1}/2,1/2)}+\mathbf{F}_{(1/2,1/2,\bar{1}/2)} \\
+&\mathbf{F}_{(1/2,\bar{1}/2,\bar{1}/2)}+\mathbf{F}_{(\bar{1}/2,1/2,\bar{1}/2)}+\mathbf{F}_{(\bar{1}/2,\bar{1}/2,1/2)}+\mathbf{F}_{(\bar{1}/2,\bar{1}/2,\bar{1}/2)} \big]. \label{divergence_111}
\end{split}
\]
One of the $\langle 111 \rangle$ fluxes is evaluated by
\[
\mathbf{F}_{(1/2,1/2,1/2)}=\bar{\alpha}_{(1/2,1/2,1/2)}\frac{\left[\beta_{(1,1,1)}-\beta_{(0,0,0)}\right]}{\sqrt{3}h},
\]
where the value of $\bar{\alpha}_{(1/2,1/2,1/2)}$ is taken by averaging the eight nearest lattice points, i.e.,
\[
\begin{split}
\bar{\alpha}_{(1/2,1/2,1/2)}=\frac{1}{8} \big[& \alpha_{(0,0,0)}+\alpha_{(1,0,0)}+\alpha_{(0,1,0)}+\alpha_{(0,0,1)} + \alpha_{(1,1,0)}+\alpha_{(1,0,1)}+\alpha_{(0,1,1)}+\alpha_{(1,1,1)} \big].
\end{split}
\]
Similarly, the other seven $\langle 111 \rangle$ fluxes in Eq. \eqref{divergence_111} are evaluated in this manner.

\section{Analyses and results in 3D} \label{Sec:3D_2}
The anisotropy of a 3D discretization can be examined either in the real space or in the Fourier space. However, there are three rotational degrees of freedom in 3D, while only one in 2D. Examining the anisotropy of a discretization in the real space in 3D is considerably more complex than that in 2D. Hence, in this section, we examine discretizations of Laplacian and divergence only in the 3D Fourier space, and then implement them for 3D PF simulations.

\subsection{Fourier space}
\subsubsection{Laplacian} \label{Sec:3D_Lap_Fourier}
The discrete Fourier transform of a 3D Laplacian follows the same definition as in Eq. \eqref{Fourier_Laplacian}. For 3D Laplacian in the continuum limit, the Fourier transform gives $\mathcal{L}(\mathbf{k})=-k^2$. For base discretizations in Eqs. \eqref{3D_L100}-\eqref{3D_L111}, the Fourier transform in the small wave number limit gives
\[
\mathcal{L}_{1,0,0} (\mathbf{k})=-k^2+\frac{1}{12}(k^4_x+k^4_y+k^4_z)+\mathcal{O}(k^6_\alpha),
\]
\[
\mathcal{L}_{0,1,0} (\mathbf{k})=-k^2+\frac{1}{12}(k^4_x+k^4_y+k^4_z+3k^2_x k^2_y+3k^2_x k^2_z+3k^2_y k^2_z)+\mathcal{O}(k^6_\alpha),
\]
\[
\mathcal{L}_{0,0,1} (\mathbf{k})=-k^2+\frac{1}{12}(k^4_x+k^4_y+k^4_z+6k^2_x k^2_y+6k^2_x k^2_z+6k^2_y k^2_z)+\mathcal{O}(k^6_\alpha).
\]
Those three base discretizations are anisotropic because their leading errors at order $k^4$ are not invariant under coordinate transformation.

\begin{figure}[h]
\centering
\includegraphics[scale=0.65]{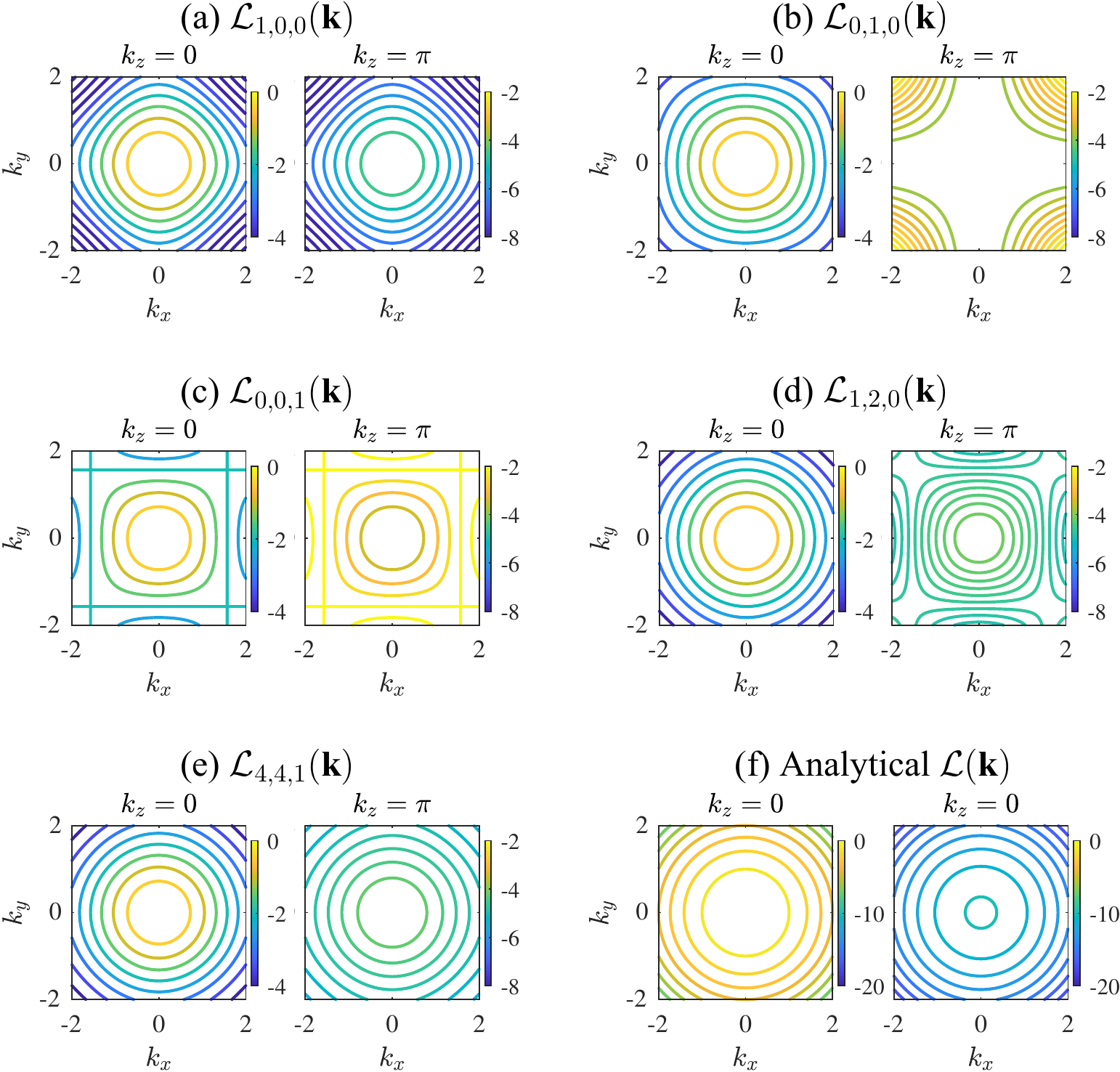}
\caption{Fourier transform of the Laplacian for different discretizations (a)-(e), and for the analytical Laplacian in the continuum limit (f).
\label{fig:Fourier_3D_Lap}}
\end{figure}

The linear combination $(c_1\mathcal{L}_{1,0,0}+c_2\mathcal{L}_{0,1,0}+c_3\mathcal{L}_{0,0,1})$ gives a complete set of discretizations involving $\langle 100 \rangle$, $\langle 110 \rangle$, and $\langle 111 \rangle$ lattice shells. There are two constraints for an isotropic discretization:
\[
c_1+c_2+c_3=1, \label{Constrain_Lap1}
\]
and
\[
3c_2+6c_3=2. \label{Constrain_Lap2}
\]
The first constraint in Eq. \eqref{Constrain_Lap1} ensures the sum of weights on three bases equals to 1. The second constraint in Eq. \eqref{Constrain_Lap2} ensures the leading error of the linear combination proportional to $k^4=(k^2_x+k^2_y+k^2_z)^2$. Since there are only two constraints to solve for three weight factors, there exist infinite solutions for an isotropic Laplacian. Among them, only three solutions put a zero weight on one base, including $\mathcal{L}_{1,2,0}$, $\mathcal{L}_{2,0,1}$ and $\mathcal{L}_{0,4,\bar{1}}$. Here $\mathcal{L}_{1,2,0}$, or the 18-point formula, is the only isotropic discretization that utilizes the first two lattice shells in the $\left<100\right>$ and $\left<110\right>$ directions. The remaining solutions of Eqs. \eqref{Constrain_Lap1}-\eqref{Constrain_Lap2} put nonzero weights on all three bases. Some individual solutions of the isotropic Laplacian have been introduced in previous studies, such as $\mathcal{L}_{5,6,1}$ in Ref. \cite{Kumar2004}, $\mathcal{L}_{7,6,2}$ in Ref. \cite{Patra2006}, and $\mathcal{L}_{4,4,1}$ (D3Q27) in Ref. \cite{Thampi2013}. The Fourier transforms of some discretizations are shown in Fig. \ref{fig:Fourier_3D_Lap}, where cross sections of 3D isocontours are taken at $k_z=0$ and $k_z=\pi$ in the $k$-space. Note that $\mathcal{L}_{1,2,0}$ is isotropic only in the cross section $k_z=0$, and $\mathcal{L}_{4,4,1}$ is isotropic in both cross sections.

\subsubsection{Divergence}
The discrete Fourier transform of the 3D divergence follows the same definition as in Eq. \eqref{Fourier_divergence}. For divergence in the continuum limit, the Fourier transform gives $\widetilde{\mathcal{D}}(\mathbf{k})=-2k^2$. For base discretizations in Eqs. \eqref{divergence_100}-\eqref{divergence_111}, the Fourier transform in the small wave number limit gives
\[
\widetilde{\mathcal{D}}_{1,0,0}(\mathbf{k})=-2k^2+\frac{2}{3}\left(k_x^4+k_y^4+k_z^4+\frac{3}{4}k^2_x k^2_y+\frac{3}{4}k^2_x k^2_z+\frac{3}{4}k^2_y k^2_z \right) +\mathcal{O}(k^6_\alpha),
\]
\[
\widetilde{\mathcal{D}}_{0,1,0}(\mathbf{k})=-2k^2+\frac{2}{3}\left(k_x^4+k_y^4+k_z^4+\frac{21}{8}k^2_x k^2_y+\frac{21}{8}k^2_x k^2_z+\frac{21}{8}k^2_y k^2_z \right) +\mathcal{O}(k^6_\alpha),
\]
\[
\widetilde{\mathcal{D}}_{0,0,1}(\mathbf{k})=-2k^2+\frac{2}{3}\left(k_x^4+k_y^4+k_z^4+\frac{9}{2}k^2_x k^2_y+\frac{9}{2}k^2_x k^2_z+\frac{9}{2}k^2_y k^2_z \right) +\mathcal{O}(k^6_\alpha).
\]
Those three base discretizations are anisotropic because their leading errors at order $k^4$ are not invariant under coordinate transformation.

\begin{figure}[h]
\centering
\includegraphics[scale=0.65]{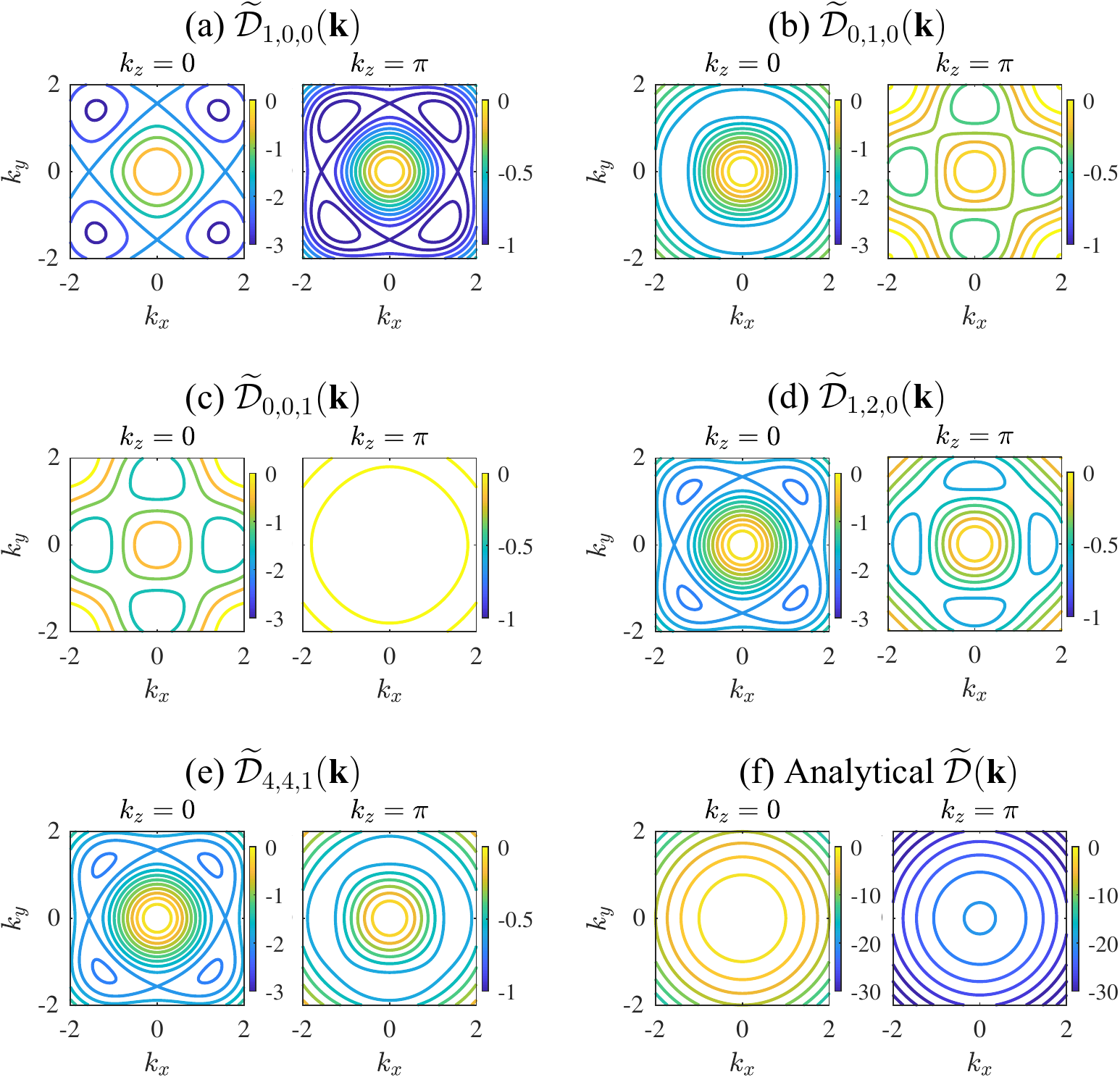}
\caption{Fourier transform of the divergence for different discretizations (a)-(e), and for the analytical divergence in the continuum limit (f). \label{fig:Fourier_3D_Div} }
\end{figure}

The linear combination $(c_1\widetilde{\mathcal{D}}_{1,0,0}+c_2\widetilde{\mathcal{D}}_{0,1,0}+c_3\widetilde{\mathcal{D}}_{0,0,1})$ gives a complete set of discretizations involving $\langle 100 \rangle$, $\langle 110 \rangle$, and $\langle 111 \rangle$ lattice shells. There are two constraints for an isotropic discretization:
\[
a_1+a_2+a_3=1, \label{Constrain_Div1}
\]
and
\[
\frac{3}{4}a_1+ \frac{21}{8}a_2+ \frac{9}{2}a_3=2. \label{Constrain_Div2}
\]
Similar to the Laplacian, the first constraint ensures the sum of weights on three bases equals to 1. The second constraint in Eq. \eqref{Constrain_Lap2} ensures the leading error of the linear combination proportional to $k^4$. Since there are only two constraints to solve for three weight factors, there exist infinite solutions for an isotropic divergence. Among them, only three solutions put a zero weight on one base, including $\widetilde{\mathcal{D}}_{1,2,0}$, $\widetilde{\mathcal{D}}_{2,0,1}$, and $\widetilde{\mathcal{D}}_{0,4,\bar{1}}$. It is worth noting that, although the constraint in Eq. \eqref{Constrain_Div2} is different from the constraint in Eq. \eqref{Constrain_Lap2} for deriving the isotropic Laplacian, they yield the same solutions. The remaining solutions of Eqs. \eqref{Constrain_Div1}-\eqref{Constrain_Div2} put nonzero weights on all three bases, such as $\widetilde{\mathcal{D}}_{5,6,1}$, $\widetilde{\mathcal{D}}_{7,6,2}$, and $\widetilde{\mathcal{D}}_{4,4,1}$. The Fourier transforms of some discretizations are shown in Fig. \ref{fig:Fourier_3D_Div}.

\subsection{Phase-field simulations in 3D}

\subsubsection{Numerical implementation}
In this section, we aim at characterizing effects of finite-difference schemes on 3D PF simulations of crystal growth. Similar to notations in 2D, we use $\mathcal{S}_{s_1,s_2,s_3}$ to denote a finite-difference scheme with the standard anti-trapping $\mathcal{A}$ in Eq. \eqref{Anti_convention}, and use $\bar{\mathcal{S}}_{s_1,s_2,s_3}$ to denote a scheme with the approximated anti-trapping $\bar{\mathcal{A}}$ in Eq. \eqref{Anti_approx}. The subscripts $s_1$, $s_2$, and $s_3$ indicate the ratio of base discretizations for each leading differential terms as in \eqref{tau_3D}. Note that we use the method of Refs. \cite{Tourret2017,Clarke2017} to discretize divergences of the supersaturation gradient and the anti-trapping current for the anisotropic scheme ${\mathcal{S}}_{1,0,0}$, where only the $\left<100\right>$ lattice shell is involved. In addition, since an isotropic $\mathcal{A}$ in 3D would require higher-order discretization of $|\vec{\nabla} \psi|$, which is much more complicated and inefficient than that in 2D, we consider only $\bar{\mathcal{A}}$ for isotropic discretizations in 3D. 

Although it is possible to use the third base for an isotropic scheme, such as $\bar{\mathcal{S}}_{2,0,1}$ and $\bar{\mathcal{S}}_{4,4,1}$, it may lead to numerical instabilities because of the large effective lattice spacing. For a moderate choice $h=1\,W$, the effective lattice spacing of the $\langle 111 \rangle$ lattice shell is $\sqrt{3} \, W$, which can lead to significant pinning effects in PF simulations \cite{Karma1998}. Thus, we implement only those finite-difference schemes without using the third base, including the anisotropic scheme $\mathcal{S}_{1,0,0}$ and the isotropic scheme $\bar{\mathcal{S}}_{1,2,0}$. For the latter, the isotropic discretization of the anti-trapping term $\bar{\mathcal{A}}_{1,2,0}$ has a similar form as $\widetilde{\mathcal{D}}_{1,2,0}$.

We use similar processes as in Section \ref{Sec:2D_results} to implement finite-difference schemes $\mathcal{S}_{1,0,0}$ and $\bar{\mathcal{S}}_{1,2,0}$ for PF simulations. But the processes of measuring the tip selection constant in 3D are slightly different. We still track the position of the most advanced solid-liquid interface along the $y$-direction, and acquire its location ($x_{tip}$, $y_{tip}$, $z_{tip}$). Then, we measure tip radii in both longitudinal and transverse planes. It follows four steps: firstly, we use the ParaView software to read the final output of $\psi$-field to find the contour of $\psi=0$; secondly, we use a MATLAB script to read the contour and rotate it by $(-\alpha)$ such that four arms of the dendrite in the $x$-$y$ plane are aligned to the lattice; thirdly, we extract the tip radius $\rho_x$ in the $z=z_{tip}$ plane by fitting the interface shape to a parabola $y=y_{tip}-(x-x_{tip})^2/(2 \rho_x)$, and select the radius that corresponds to a fit until a distance of one radius behind the tip location; lastly, we repeat the third step to extract the tip radius $\rho_z$ in the $x=x_{tip}$ plane, and calculate the radius of mean curvature $\rho$ with the equation $1/\rho=(1/\rho_x+1/\rho_z)/2$ \cite{Clarke2017}. The value of $\rho$ is used to evaluate the tip selection constant.

\subsubsection{Isothermal solidification}

\begin{figure}[h]
\centering
\includegraphics[scale=0.3]{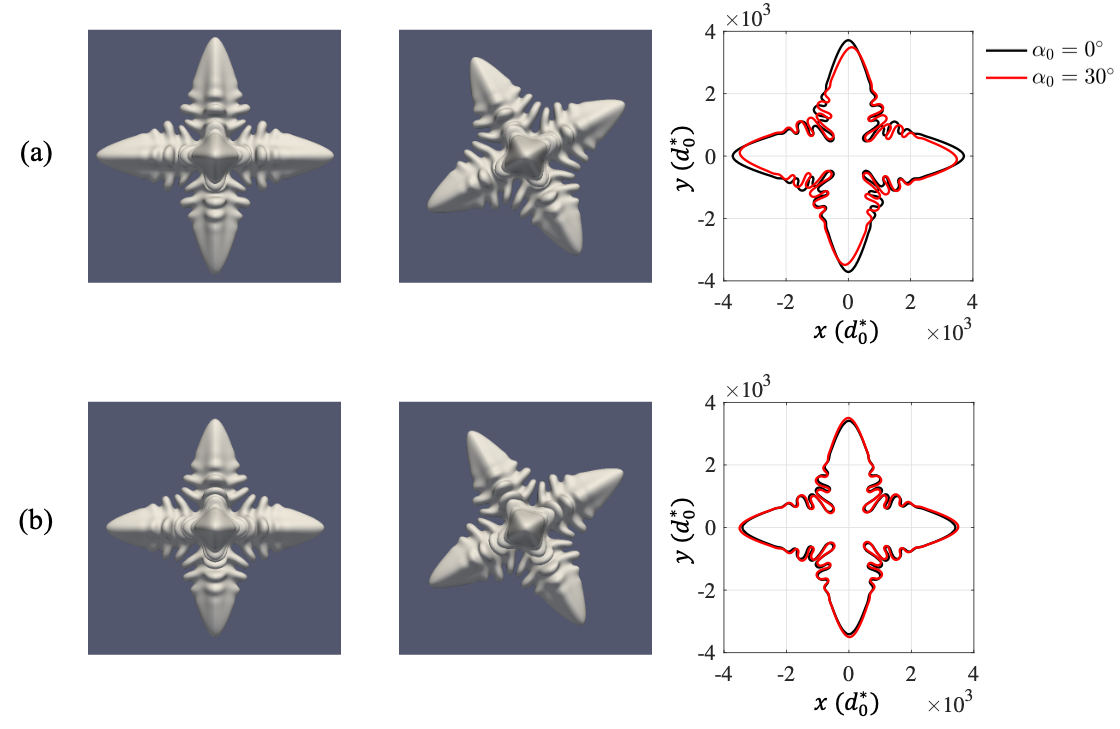}
\caption{Comparison of 3D PF simulations with different crystal orientations $\alpha_0=0^{\circ}$ and $\alpha_0=30^{\circ}$. Simulations use two finite-difference schemes: (a) the anisotropic scheme $\mathcal{S}_{1,0,0}$ and (b) the isotropic scheme $\bar{\mathcal{S}}_{1,2,0}$. The contours of $\alpha_0=30^{\circ}$ are rotated back by $(-\alpha_0)$ to be compared with the contours of $\alpha_0=0^{\circ}$. \label{fig:3D_examples}}
\end{figure}

Initially, a spherical seed of radius $r = 50$ $(d_0^*)$ is placed at the center of a supersaturated melt with $\Omega=0.25$. The no-flux conditions are applied at boundaries
of a cubic simulation domain with the side length $9.6 \times 10^4$ $(d^*_0)$. The simulation time is set to $t=1.5 \times 10^6$ $({d_0^*}^2/D)$, with a time step $\Delta t=0.8h^2/(6D)$. If not specifically mentioned, we use parameters $h/W=1.2$, $W/d_0^*=16$, and $\epsilon_4=0.014$. PF simulations with different crystal orientations between $\alpha_0=0^{\circ}$ and $\alpha_0=30^{\circ}$ are compared in Fig. \ref{fig:3D_examples}. The superposed solid-liquid interfaces show that, while the anisotropic scheme $\mathcal{S}_{1,0,0}$ breaks the rotational invariance, the isotropic scheme $\bar{\mathcal{S}}_{1,2,0}$ is able to maintain it.

\begin{figure}[h]
\centering
\includegraphics[scale=0.65]{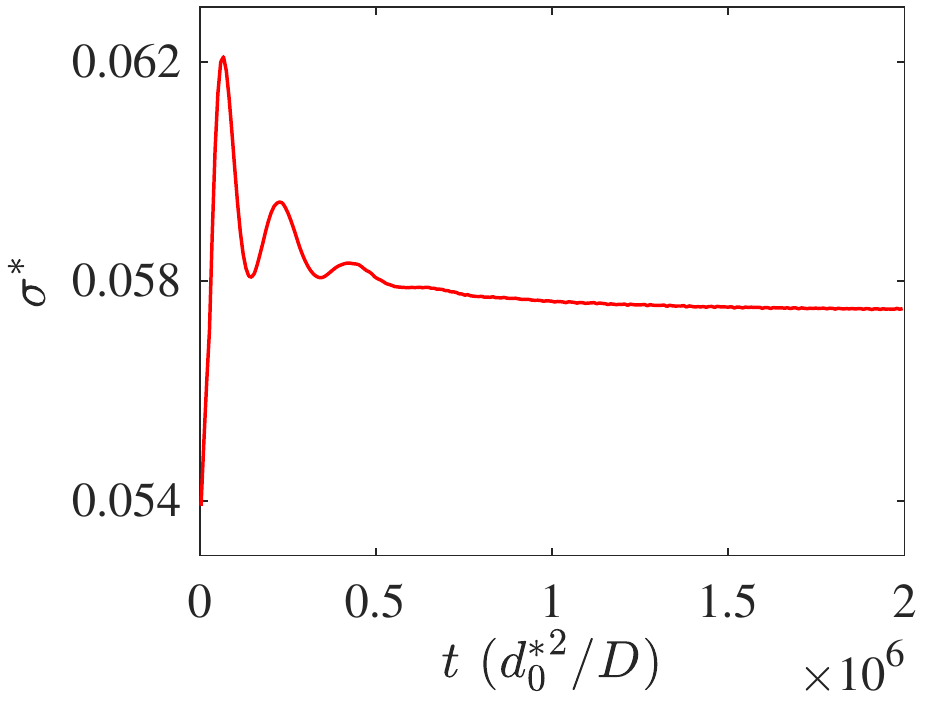}
\caption{
The tip selection constant $\sigma^*$ as a function of time in a 3D PF simulation using the $\mathcal{S}_{1,0,0}$ scheme, with $\alpha_0=0^{\circ}$, $W/d_0^*=8$, and $h/W= 1.2$.
\label{fig:3D_sigma_t}}
\end{figure}

\begin{figure}[h]
\centering
\includegraphics[scale=0.65]{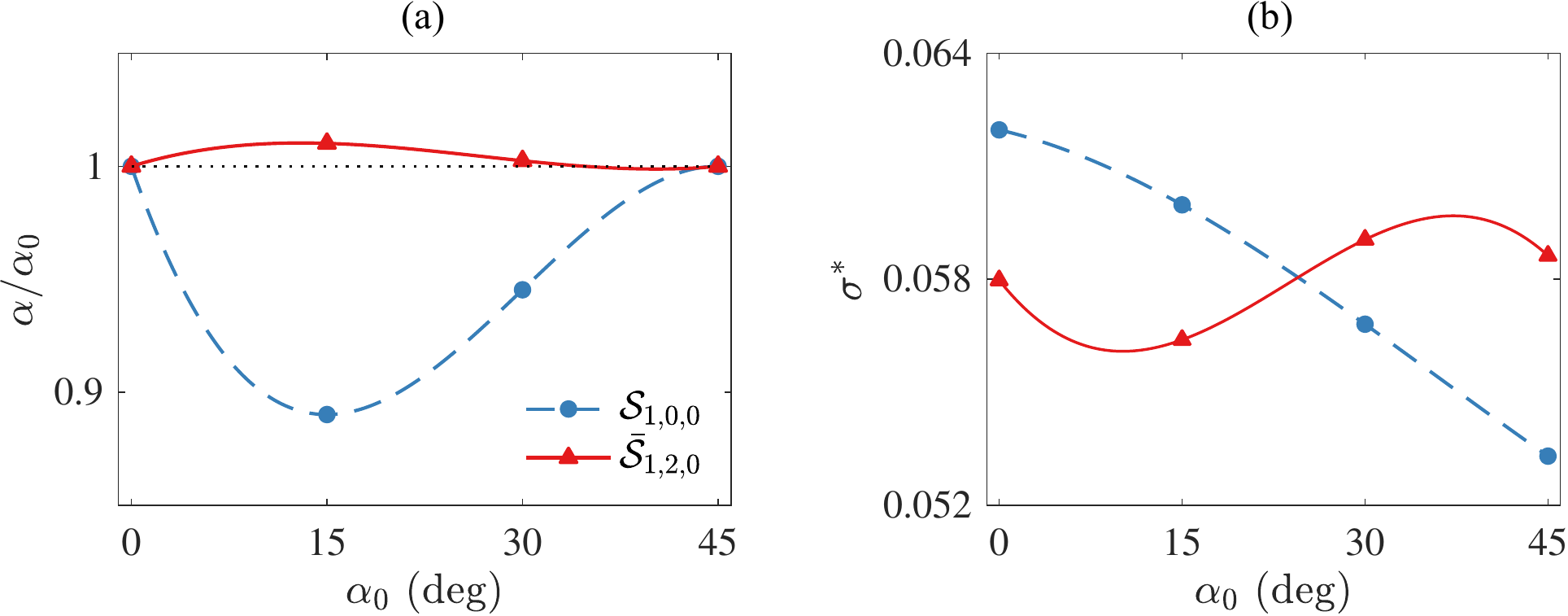}
\caption{Measurements of $\alpha/\alpha_0$ and $\sigma^*$ as a function of $\alpha_0$ for PF simulations of isothermal crystal growth. \label{fig:3D_angles}}
\end{figure}

We also use $\alpha/\alpha_0$ and $\sigma^*$ as two criteria to examine effects of finite-difference schemes in 3D. The time-dependence of $\sigma^*$ in a 3D PF simulation with the scheme $\mathcal{S}_{1,0,0}$ and $\alpha_0=0^{\circ}$ is shown in Fig. \ref{fig:2D_sigma_t}, where we measured $\sigma^*$ at $t=1.5 \times 10^6$ $({d_0^*}^2/D)$. We measured $\sigma^*$ at the same time when it reached steady-state in the other 3D simulations with different schemes and crystal orientations. Fig. \ref{fig:3D_angles} shows $\alpha/\alpha_0$ and $\sigma^*$ as a function of $\alpha_0$. It is clear that the isotropic scheme $\bar{\mathcal{S}}_{1,2,0}$ improves the rotational invariance in terms of both criteria for isothermal crystal growth in 3D PF simulations.

The performance of finite-difference schemes in 3D is summarized in Table \ref{tab:3D_performance}, where we measured the efficiency of PF simulations in Fig. \ref{fig:3D_angles}. Because the isotropic scheme $\bar{\mathcal{S}}_{1,2,0}$ requires more computation, it is less efficient than the anisotropic scheme $\mathcal{S}_{1,0,0}$.

\begin{table}[h]
\caption{The performance of three finite-different schemes in 3D. The efficiency measures the time a PF simulation takes on an NVIDIA Tesla P100 GPU. \label{tab:3D_performance}}
\centering
\begin{tabular}{cccc} \hline
Scheme & Anti-trapping & Property & Efficiency ($s$)\\\hline
$\mathcal{S}_{1,0,0}$ & ${\mathcal{A}}(\mathbf{r})$ & Anisotropic & 5445 \\
$\bar{\mathcal{S}}_{1,2,0}$ & $\bar{\mathcal{A}}(\mathbf{r})$ & Isotropic & 25155 \\
\hline
\end{tabular}
\end{table}

\subsubsection{Directional solidification}
In subsection \ref{sec:directional_2D} for 2D PF simulations of directional solidification, we have shown that the PF simulations with isotropic finite-difference implementation can yield a characteristic length $\delta$ that agrees well with the analytical solution. In this subsection, we perform 3D PF simulations of directional solidification, and focus on the influence of 3D finite-difference implementation on the tip selection constant $\sigma$, defined as $\sigma=2 D d_0/\rho^2 V_p$ at the reference temperature $T=T_0$ (section \ref{Sec:evolution_eqs}). For the temperature gradient considered here ($G=19$ $\mathrm{K/cm}$), the local temperature variation around the tip is negligible. Thus, the dendritic tip growth can be considered as locally isothermal at the steady-state, and the presence of $G$ does not lead to noticeable asymmetry in the tip region. In addition, since $\sigma$ only depends on the anisotropy strength according to the solvability theory \cite{Barbieri1989}, $\sigma$ should not vary significantly as a function of $\alpha_0$ when diffusion fields have a negligible overlap between dendritic tips. Therefore, using $\sigma$ to examine the rotational invariance is still valid for PF simulations of directional solidification.

\begin{figure}[h]
\centering
\includegraphics[scale=0.3]{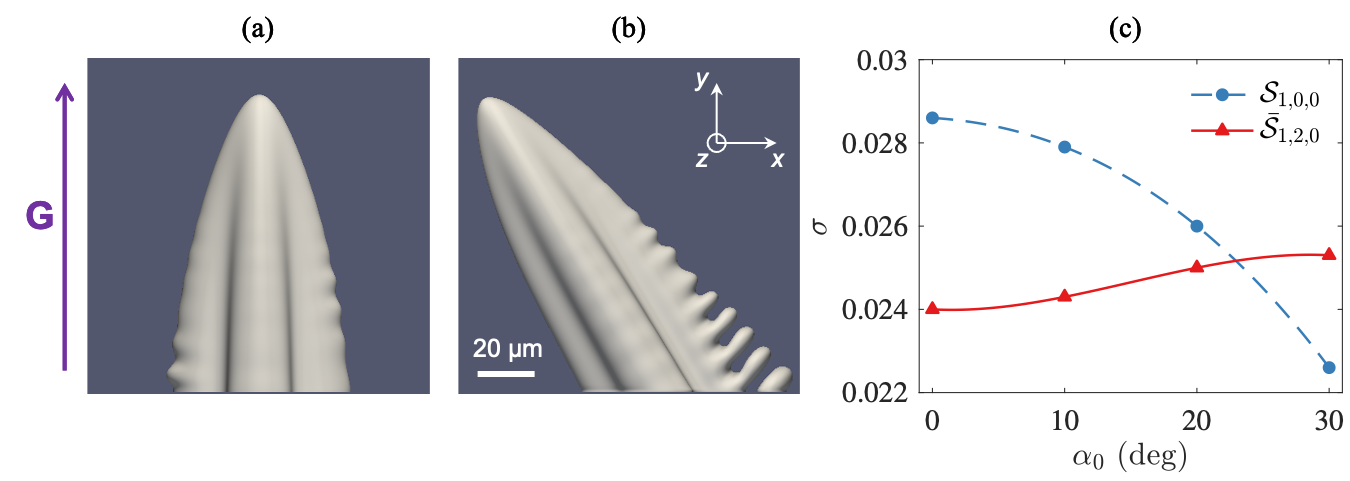}
\caption{3D PF simulations of directional solidification. The isotropic finite-difference scheme $\bar{\mathcal{S}}_{1,2,0}$ is used for simulations with misorientation angles $\alpha_0$ between the preferred $\left<100\right>$ growth direction and the temperature gradient: (a) $\alpha_0=0^{\circ}$ and (b) $\alpha_0=30^{\circ}$. (c) The tip selection constant $\sigma$ as a function of $\alpha_0$. Compared with the anisotropic scheme $\mathcal{S}_{1,0,0}$, the isotropic scheme $\bar{\mathcal{S}}_{1,2,0}$ can significantly improve the rotational invariance in terms of $\sigma$.
\label{fig:3D_DS}}
\end{figure}

Results of 3D PF simulations for directional solidification of a SCN-1.3 wt.\% acetone alloy with $V_p=86$ $\mathrm{\mu m/s}$ and $G=19$ $\mathrm{K/cm}$ are shown in Fig. \ref{fig:3D_DS}. The simulation and material parameters used in Fig. \ref{fig:3D_DS} include: the interface thickness $W/d_0 = 194.2$, the grid spacing $h/W = 1$, the explicit time step $t=1.46 \times 10^{-5}$ s, the crystalline anisotropy $\epsilon_4=0.007$, the partition coefficient $k=0.1$, the liquidus slope $m = -3.02$ $\mathrm{K/wt.\%}$, and the diffusion coefficient $D = 1270$ $\mathrm{\mu m^2/s}$. The simulation domain size is $L_x\times L_y\times L_z= 101\times203\times76$ $\mathrm{\mu m^3}$. Periodic boundary conditions are applied in the $x$ directions, and wetting boundary condition \cite{Tourret2017,Clarke2017} are applied in the $z$ directions.

As shown in Fig. \ref{fig:3D_DS}(c), there is a significant variation of $\sigma$ as a function of $\alpha_0$ for the anisotropic scheme ${\mathcal{S}}_{1,0,0}$, which brings considerable spurious effects to PF simulations. In comparison, the isotropic scheme $\bar{\mathcal{S}}_{1,2,0}$ has led to notable improvement of the rotational invariance and there is no significant variation of $\sigma$ as a function of $\alpha_0$. Under such a circumstance, using the isotropic scheme is necessary for a quantitative PF study. However, PF simulations using the $\bar{\mathcal{S}}_{1,2,0}$ scheme take about five times longer than using the conventional ${\mathcal{S}}_{1,0,0}$ scheme, as shown in Table \ref{tab:3D_performance}. Thus, there should be trade-offs when adopting this more accurate but less efficient finite-difference scheme in 3D.

Within a finite-difference scheme, the isotropic discretization of the anti-trapping has the critical influence on PF simulations, whereas the isotropic discretizations of other leading differential terms play more secondary roles. The spurious effects of the lattice anisotropy are found to be significant for large diffusive interface thickness, large grid spacing, large misorientations, and/or high growth velocity when the amount of the compensating anti-trapping current is large. In these scenarios, it is necessary to use the rotationally invariant scheme to avoid spurious effects. However, using an isotropic scheme in 3D is computationally more demanding (Table \ref{tab:3D_performance}). Therefore, in certain applications, such as PF simulations of directional solidification in the presence of subgrain boundaries where the misorientation angle with respect to the thermal axis is typically only a few degrees \cite{Mota2021}, the use of a  
standard discretization can be computationally more efficient while remaining equally accurate (Fig. \ref{fig:3D_DS}(c)).

\section*{Conclusion}
We systematically examined the rotational invariance of several finite-difference schemes for PF models in 2D and 3D. A rotationally invariant scheme requires isotropic discretizations of leading differential terms in equations of motion, including $\mathcal{L}(\mathbf{r})$, $\mathcal{G}(\mathbf{r})$, $\mathcal{D}(\mathbf{r})$, and the divergence of the anti-trapping current. Since the amount of the compensating anti-trapping current increases with the velocity and thickness of a diffuse interface in PF simulations, the isotropic discretization of the anti-trapping term is especially important. We used known methods in the real or Fourier space to derive finite-difference approximations of those leading differential terms in 2D and 3D that are isotropic at order $h^2$ of the lattice spacing $h$. An approximated form of the anti-trapping current was proposed based on the standard tangent hyperbolic profile of a stationary PF at an equilibrium interface, which makes its form similar to a generalized divergence and facilitates the Fourier-space derivation of the associated isotropic discretizations in both 2D and 3D. We also derived an isotropic discretization of the standard anti-trapping term in 2D, which requires a larger stencil for a higher-order discretization of the gradient and becomes more complex and inefficient in 3D.

In addition, we proposed a general framework for identifying isotropic discretizations through the linear combination of base discretizations that utilize lattice shells containing a group of lattice points in each set of equivalent directions. In 2D, there are two base discretizations that utilize lattice shells in the $\left<10\right>$ and $\left<11\right>$ directions, respectively. A unique isotropic discretization can be obtained through the linear combination of those two bases, with $2/3$ weight on the $\left<10\right>$ base and $1/3$ weight on the $\left<11\right>$ base. In 3D, there are three base discretizations that utilize lattice shells in the $\left<100\right>$, $\left<110\right>$ and $\left<111\right>$ directions, respectively. We can obtain various isotropic discretizations through the linear combination of those three bases. Since the $\left<111\right>$ base that can cause pinning effect in PF simulations should be avoided, we derived 3D isotropic discretizations involving only the first two bases, with $1/3$ weight on the $\left<100\right>$ base and $2/3$ weight on the $\left<110\right>$ base. These conclusions apply to all leading differential terms in 2D and 3D.

We implemented isotropic finite-difference schemes for PF simulations of isothermal and directional solidification in both 2D and 3D. For isothermal crystal growth, those isotropic schemes can dramatically reduce spurious lattice anisotropy effects and significantly improve the convergence of PF simulations in terms of both the crystal growth orientation and the tip selection constant. A comparison of 2D PF simulations using the approximated anti-trapping $\tilde{\mathcal{A}}_{2,1}$ and the full anti-trapping ${\mathcal{A}}_{2,1}$ also shows that this approximation is quantitatively accurate and turns out to be modestly more computationally efficient. For dendritic crystal growth during directional solidification in 2D, only those isotropic finite-difference schemes can accurately resolve the dendritic tip and yield the characteristic length scale $\delta$ that agrees well with the analytical solution, which is essential for the accurate modeling of grain competition in the well-developed dendritic regime.

The isotropic finite-difference schemes proposed in this paper can potentially be applied to PF simulations of rapid solidification for fusion-based additive manufacturing of metals \cite{Francois2017,Ghosh2018}, where the solid-liquid interface is far from equilibrium and grains are polycrystalline. Furthermore, they can also be applied to other numerical problems that require isotropic discretizations of Laplacian, divergence, and gradient terms.

\section*{Acknowledgement}
This research was supported by NASA under award No. 80NSSC19K0135 in the framework of the CETSOL project. The majority of numerical simulations were performed on the Discovery cluster of Northeastern University located in Massachusetts Green High Performance Computing Center (MGHPCC) in Holyoke, MA.

\appendix

\section{Isotropic discretization of gradient square} \label{Sec:dis_grad}



Consider the gradient square term $\mathcal{G}(\mathbf{r})=|\vec{\nabla} \psi|^2$. In 2D, two base discretizations of $\mathcal{G}(\mathbf{r})$ are 
\[
\mathcal{G}_{1,0}=\frac{1}{4 h^2 } \left[  (\psi_{(1,0}-\psi_{(\bar{1},0)})^2+(\psi_{(0,1)}-\psi_{(0,\bar{1})})^2  \right], \label{Grad_10}
\]
and
\[
\mathcal{G}_{0,1}=\frac{1}{8 h^2 } \left[  (\psi_{(1,1}-\psi_{(\bar{1},\bar{1})})^2+(\psi_{(1,\bar{1})}-\psi_{(\bar{1},1)})^2  \right]. \label{Grad_11}
\]
The discrete Fourier transform of $\mathcal{G}(\mathbf{r})$ is defined as
\[
\mathcal{G} (\mathbf{k})=\frac{\sum_{\mathbf{r}} e^{-i\mathbf{k}\cdot\mathbf{r}} \mathcal{G}(\mathbf{r}) } {\sum_{\mathbf{r}} e^{-i\mathbf{k} \cdot\mathbf{r}}\psi(\mathbf{r})\psi(\mathbf{r})}. \label{Fourier_Gradient}
\]
It is normalized by the convolution of the same function in the Fourier space, i.e., $\mathcal{F}[\psi(\mathbf{r})\psi(\mathbf{r})]=\mathcal{F}[\psi(\mathbf{r})]*\mathcal{F}[\psi(\mathbf{r})]$, where $\mathcal{F}[\psi(\mathbf{r})]$ is the Fourier transform of the scalar field $\psi(\mathbf{r})$, i.e., $\mathcal{F}[\psi(\mathbf{r})]=\sum_{\mathbf{r}} e^{-i\mathbf{k} \cdot\mathbf{r}}\psi(\mathbf{r})$. For gradient in the continuum limit, the Fourier transform gives $\mathcal{G} (\mathbf{k})=-k^2$. For two bases in Eqs. \eqref{Grad_10} and \eqref{Grad_11}, it gives
\[
\mathcal{G}_{1,0} (\mathbf{k})=-k^2+\frac{1}{3}(k^4_x+k^4_y)+\mathcal{O}(k^6_\alpha),
\]
and
\[
\mathcal{G}_{0,1} (\mathbf{k})=-k^2+\frac{1}{3}(k^4_x+6k^2_xk^2_y+k^4_y)+\mathcal{O}(k^6_\alpha).
\]
Similar to the Laplacian and the divergence, one can show that $\mathcal{G}_{2,1}$ is the unique isotropic discretization of the gradient through linear combination of $\mathcal{G}_{1,0}$ and $\mathcal{G}_{0,1}$.

In 3D, three base discretizations of $\mathcal{G}(\mathbf{r})$ are
\[
\mathcal{G}_{1,0,0}=\frac{1}{4 h^2 } \left[  (\psi_{(1,0,0}-\psi_{(\bar{1},0,0)})^2+(\psi_{(0,1,0)}-\psi_{(0,\bar{1},0)})^2+(\psi_{(0,0,1)}-\psi_{(0,0,\bar{1})})^2  \right], \label{Grad_100}
\]
\[
\begin{split}
\mathcal{G}_{0,1,0}=\frac{1}{16 h^2 } \big[  &(\psi_{(1,1,0}-\psi_{(\bar{1},\bar{1},0)})^2+(\psi_{(1,\bar{1},0)}-\psi_{(\bar{1},1,0)})^2+(\psi_{(1,0,1)}-\psi_{(\bar{1},0,\bar{1})})^2 \\
+&(\psi_{(\bar{1},0,1)}-\psi_{(1,0,\bar{1}})^2+(\psi_{(0,1,1)}-\psi_{(0,\bar{1},\bar{1})})^2+(\psi_{(0,\bar{1},1)}-\psi_{(0,1,\bar{1})})^2 \big], \label{Grad_110}
\end{split}{}
\]
and
\[
\begin{split}
\mathcal{G}_{0,0,1}=\frac{1}{144 h^2 } \big[  &(\psi_{(1,1,1}-\psi_{(\bar{1},\bar{1},\bar{1})})^2+(\psi_{(1,1,\bar{1}}-\psi_{(\bar{1},\bar{1},1})^2 \\
+&(\psi_{(1,\bar{1},1}-\psi_{(\bar{1},1,\bar{1})})^2+(\psi_{(\bar{1},1,1}-\psi_{(1,\bar{1},\bar{1})})^2 \big].
\end{split} \label{Grad_111}
\]
Following Eq. \eqref{Fourier_Gradient}, the discrete Fourier transform of those three base discretizations yields
\[
\mathcal{G}_{1,0,0} (\mathbf{k})=-k^2+\frac{1}{3}(k^4_x+k^4_y+k^4_z)+\mathcal{O}(k^6_\alpha),
\]
\[
\mathcal{G}_{0,1,0} (\mathbf{k})=-k^2+\frac{1}{3}(k^4_x+k^4_y+k^4_z+3k^2_xk^2_y+3k^2_xk^2_z+3k^2_yk^2_z)+\mathcal{O}(k^6_\alpha),
\]
\[
\mathcal{G}_{0,0,1} (\mathbf{k})=-k^2+\frac{1}{3}(k^4_x+k^4_y+k^4_z+6k^2_xk^2_y+6k^2_xk^2_z+6k^2_yk^2_z)+\mathcal{O}(k^6_\alpha).
\]
The linear combination $(c_1\mathcal{G}_{1,0,0}+c_2\mathcal{G}_{0,1,0}+c_3\mathcal{G}_{0,0,1})$ gives a complete set of discretizations involving the $\langle 100 \rangle$, $\langle 110 \rangle$, and $\langle 111 \rangle$ lattice shells. There are two constraints for an isotropic discretization:
\[
c_1+c_2+c_3=1,
\]
and
\[
3c_2+6c_3=2.
\]
The first constraint ensures the sum of weights on three bases equals to 1. The second constraint ensures the leading error of the linear combination proportional to $k^4$. Since there are only two constraints to solve for three weight factors, there exist infinite solutions for an isotropic gradient. Among them, only three solutions put a zero weight on one base, including $\mathcal{G}_{1,2,0}$, $\mathcal{G}_{2,0,1}$, and $\mathcal{G}_{0,4,\bar{1}}$. The remaining solutions put nonzero weights on all three bases, such as $\mathcal{G}_{5,6,1}$, $\mathcal{G}_{7,6,2}$, and $\mathcal{G}_{4,4,1}$.

\section{Isotropic discretization of full anti-trapping} \label{Sec:appendix_vanish}
Consider a generalized form of the full anti-trapping ${\mathcal{A}}(\mathbf{r})=\vec{\nabla} \cdot \widetilde{\mathbf{F}}=\vec{\nabla}\cdot( \alpha {\vec{\nabla}\beta}/{|\vec{\nabla}\beta|} )$, where $\alpha=[(1+(1-k) U) (1-\varphi^2) \partial_t \psi]/2$ and $\beta=\psi$ are two scalars. When the discretization of $|\vec{\nabla}\beta|$ has vanishing error at order $h^2$, it can be absorbed into the flux that is used for discretizing $\widetilde{\mathcal{D}}(\mathbf{r})$, where its higher-order errors will not contribute to the leading error of a discretized ${\mathcal{A}}(\mathbf{r})$ at order $h^2$. Similar to $\widetilde{\mathcal{D}}_{2,1}$, the isotropic discretization ${\mathcal{A}}_{2,1}$ utilizes eight fluxes in total, including four in the $\left<10 \right>$ directions and four in $\left<11 \right>$ directions. We checked the rotational invariance of ${\mathcal{A}}_{2,1}$ in the real space using the Mathematica software and verified that ${\mathcal{A}}_{2,1}$ is an isotropic discretization. A discretized $|\vec{\nabla}\beta|$ in 2D with vanishing second-order error requires an extended stencil that contains 21 lattice points, including four in $\left<10\right>$ directions, four in $\left<11\right>$ directions, four in $\left<20\right>$ directions, eight in $\left<21\right>$ directions, and the origin point. However, a discretized $|\vec{\nabla}\beta|$ in 3D with vanishing second-order error becomes more complex and inefficient for numerical implementation. Here we only discuss the case in 2D.

For evaluating $|\vec{\nabla}\beta|$ in the denominator of the flux $\widetilde{\mathbf{F}}$ at off-lattice locations, there are two scenarios to consider: the flux at an off-lattice location with two nearest lattice points, such as $\widetilde{\mathbf{F}}_{(1/2,0)}$; the flux at an off-lattice location with four nearest lattice points, such as $\widetilde{\mathbf{F}}_{(1/2,1/2)}$. In the former scenario, $|\vec{\nabla}\beta|$ is discretized by
\[
|\vec{\nabla}\beta|_{(1/2,0)} = \sqrt{ \left( \partial_x \beta|_{({1}/{2},0)} \right)^2+\left( \partial_y \beta|_{({1}/{2},0)} \right)^2} +\mathcal{O}\left(h^4 \right),
\]
where
\[
\begin{split}
\partial_x \beta|_{(1/2,0)} =& \frac{35}{24} \frac{[\beta_{(1,1)}-\beta_{(0,\bar{1})}]-[\beta_{(0,1)}-\beta_{(1,\bar{1})}]}{2h}-\frac{1}{8}\frac{[\beta_{(2,1)}-\beta_{(\bar{1},\bar{1})}]-[\beta_{(\bar{1},1)}-\beta_{(2,\bar{1})}]}{6h} \\
&- \frac{1}{3}\frac{[\beta_{(1,2)}-\beta_{(0,\bar{2})}]-[\beta_{(0,2)}-\beta_{(1,\bar{2})}]}{2h},
\end{split}
\]
and
\[
\begin{split}
\partial_y \beta|_{(1/2,0)} =& \frac{35}{24} \frac{[\beta_{(1,1)}-\beta_{(0,\bar{1})}]+[\beta_{(0,1)}-\beta_{(1,\bar{1})}]}{4h}-\frac{1}{8}\frac{[\beta_{(2,1)}-\beta_{(\bar{1},\bar{1})}]+[\beta_{(\bar{1},1)}-\beta_{(2,\bar{1})}]}{4h} \\
&- \frac{1}{3}\frac{[\beta_{(1,2)}-\beta_{(0,\bar{2})}]+[\beta_{(0,2)}-\beta_{(1,\bar{2})}]}{8h}.
\end{split}
\]
Here the derivatives $\partial_x \beta|_{(1/2,0)}$ and $\partial_y \beta|_{(1/2,0)}$ already have vanishing second-order error. At the other off-lattice locations with two nearest-neighbor lattice points, a similar discretization of $|\vec{\nabla}\beta|_{(1/2,0)}$ can be applied through the translation and/or rotation operations. In the latter scenario, $|\vec{\nabla}\beta|$ is discretized by
\[
|\vec{\nabla}\beta|_{(1/2,1/2)} = \sqrt{ \left( \partial_x \beta|_{({1}/{2},1/2)} \right)^2+\left( \partial_y \beta|_{({1}/{2},1/2)} \right)^2} +\mathcal{O}\left(h^4 \right),
\]
where
\[
\begin{split}
\partial_x \beta|_{(1/2,1/2))} =& \frac{5}{4} \frac{[\beta_{(1,1)}-\beta_{(0,0)}]-[\beta_{(0,1)}-\beta_{(1,0)}]}{2h}-\frac{1}{8}\frac{[\beta_{(2,1)}-\beta_{(\bar{1},0)}]-[\beta_{(\bar{1},1)}-\beta_{(2,0)}]}{6h} \\
&- \frac{1}{8}\frac{[\beta_{(1,2)}-\beta_{(0,\bar{1})}]-[\beta_{(0,2)}-\beta_{(1,\bar{1})}]}{2h},
\end{split}
\]
and
\[
\begin{split}
\partial_y \beta|_{(1/2,1/2))} =& \frac{5}{4} \frac{[\beta_{(1,1)}-\beta_{(0,0)}]+[\beta_{(0,1)}-\beta_{(1,0)}]}{2h}-\frac{1}{8}\frac{[\beta_{(2,1)}-\beta_{(\bar{1},0)}]+[\beta_{(\bar{1},1)}-\beta_{(2,0)}]}{2h} \\
&- \frac{1}{8}\frac{[\beta_{(1,2)}-\beta_{(0,\bar{1})}]+[\beta_{(0,2)}-\beta_{(1,\bar{1})}]}{6h}.
\end{split}
\]
Similarly, the derivatives $\partial_x \beta|_{(1/2,1/2)}$ and $\partial_y \beta|_{(1/2,1/2)}$ already have vanishing second-order error. At the other off-lattice locations with four nearest-neighbor lattice points, a similar discretization of $|\vec{\nabla}\beta|_{(1/2,1/2)}$ can be applied through the translation operation.

\bibliographystyle{elsarticle-num-names}
\bibliography{cas-refs}

\end{document}